\documentclass[11pt]{article}
\usepackage{jheppub}
\usepackage{amsfonts,ulem,bbold}
\usepackage{amsmath,amssymb, braket}

\newcommand{\gmn}{g_{\mu\nu}}
\newcommand{\fmn}{f_{\mu\nu}}
\newcommand{\bgmn}{\bar g_{\mu\nu}}

\newcommand{\pmn}{P_{\mu\nu}}

\newcommand{\rmn}{R_{\mu\nu}}
\newcommand{\Smn}{S^\mu_{~\nu}}
\newcommand{\Pmn}{P^\mu_{~\nu}}
\newcommand{\Rmn}{R^\mu_{~\nu}}
\newcommand{\Gmn}{G^\mu_{~\nu}}

\newcommand{\be}{\begin{equation}}
\newcommand{\ee}{\end{equation}}
\newcommand{\beqn}{\begin{eqnarray}}
\newcommand{\eeqn}{\end{eqnarray}}
\newcommand{\baln}{{\begin{align}}}
\newcommand{\ealn}{{\end{align}}}
\newcommand{\td}{\mathrm{d}}

\newcommand{\p}{\partial}
\newcommand{\dd}{\mathrm{d}}
\newcommand{\nn}{\nonumber}
\newcommand{\Tr}{\mathrm{Tr}}

\def\ph{\phantom}

\title{Higher Derivative Gravity and Conformal Gravity 
From Bimetric and Partially Massless Bimetric Theory}  
\author[1]{S.F.~Hassan,}
\author[1,2]{Angnis~Schmidt-May,}
\author[1,3]{Mikael~von~Strauss}
\affiliation[1]{Department of Physics \& 
        The Oskar Klein Centre,\\
        Stockholm University, AlbaNova University Centre, 
        SE-106 91 Stockholm, Sweden}
\affiliation[2]{Institut f\"ur Theoretische Physik, Eidgen\"ossische
  Technische Hochschule Z\"urich\\ 
Wolfgang-Pauli-Strasse 27, 8093 Z\"urich, Switzerland}
\affiliation[3]{UPMC-CNRS, UMR7095,
 Institut d'Astrophysique de Paris, GReCO,\\
 98bis boulevard Arago, F-75014 Paris, France.}
 
\emailAdd{fawad@fysik.su.se}
\emailAdd{angniss@itp.phys.ethz.ch}
\emailAdd{strauss@iap.fr}

\abstract{In this paper, we establish the correspondence between
 ghost-free bimetric theory and a class of higher derivative gravity
 actions, including conformal gravity and new massive gravity. We
 also characterize the relation between the respective equations of
 motion and classical solutions. We illustrate that, in this
 framework, the spin-2 ghost of higher derivative gravity at the
 linear level is an artifact of the truncation to a four-derivative
 theory. The analysis also gives a relation between the proposed
 partially massless (PM) bimetric theory and conformal gravity,
 showing, in particular, the equivalence of their equations of motion
 at the four-derivative level. For the PM bimetric theory, this provides
 further evidence for the existence of an extra gauge symmetry and
 the associated loss of a propagating mode away from de Sitter
 backgrounds. The new symmetry is an extension of Weyl invariance,
 which may suggest the candidate PM bimetric theory as a possible
 ghost-free completion of conformal gravity.}


\begin{document} 
\maketitle
\flushbottom

\section{Introduction}

In this paper, we show the correspondence between the ghost-free
bimetric theory \cite{Hassan:2011zd,Hassan:2011ea} and higher
derivative gravity, both of which have similar spectra, but only the
bimetric case is ghost-free. \mbox{In particular}, this implies a close
relation between conformal gravity and a specific bimetric theory that
has been proposed \cite{Hassan:2012gz,Hassan:2012rq} as the
sought-after nonlinear partially massless (PM) theory. In this
section, we start with a brief discussion of various theories
considered in this paper, emphasizing the features that are of
relevance here. We then present a summary of our results. In Section
2, we consider the correspondence for general bimetric parameters, and
the ghost issue is discussed in Section 3. Section~ 4 is devoted to the
PM bimetric theory and conformal gravity. Section 5 contains some
discussions. \mbox{A scalar} field example is worked out in Appendix A, and
some calculational details are relegated to Appendix B.

\subsection{A Review of the Different Theories Considered}

Below, we briefly review the relevant features of higher derivative
gravity, conformal gravity, ghost-free bimetric theory and partially
massless theories.

\subsubsection{Higher Derivative Gravity}

By this, we mean theories with more than two derivatives of the metric
$\gmn$ that, at the four-derivative level, have the form,
\begin{align}
S^{\mathrm{HD}}_{(2)}[g]=m_g^{2}\int\dd^4x\sqrt{g}\left[\Lambda+c_R R(g)
-\frac{c_{RR}}{m^2}\left(R^{\mu\nu}R_{\mu\nu}-\frac{1}{3}R^2\right)
\right] \,.
\label{SHD2}
\end{align}

This action propagates a massless spin-2 state with two helicities,
along with a massive spin-2 state with five helicities, for a total of
seven modes \cite{Stelle:1976gc,Stelle:1977ry}. The massive spin-2
state of mass $m_2^2=m^2 c_R/c_{RR}$ is a ghost; hence, it violates
unitarity. Changing the coefficient of the $R^2$ term away from
$\frac{1}{3}$ adds a massive scalar to the spectrum. Precisely for the
form Equation \eqref{SHD2}, the scalar becomes infinitely massive and decouples.
To Equation \eqref{SHD2}, one may add higher derivative terms suppressed by
higher powers of $1/m^2$. The above action can be generalized to $d$
dimensions,
\begin{align}
S^{\mathrm{HD}}_{(2)}[g]=m_g^{d-2}\int\dd^dx\sqrt{g}\left[\Lambda+c_R
 R(g) -\frac{c_{RR}}{m^2}\left(R^{\mu\nu}R_{\mu\nu}-\frac{d}{4(d-1)}
 R^2\right) \right] \,.
\label{SHD2d}
\end{align}

For $d=3$ and $m_g\,c_R<0$, the action Equation \eqref{SHD2d} has been interpreted
as a ghost-free theory of massive gravity, dubbed new massive gravity
\cite{Bergshoeff:2009hq}. The choice $m_g\,c_R<0$ renders the massive
spin-2 state healthy. However, the massless state, which is now a ghost,
does not propagate in $d=3$, so the spectrum consists of a single
massive spin-2 state. This construction is peculiar to $d=3$ and
cannot cure the ghost problem for $d\geq 4$
\cite{Ohta:2011rv, Kleinschmidt:2012rs}.
For some work on higher derivative gravity, see
\cite{Eliezer:1989cr, Simon:1990ic, Biswas:2005qr, Biswas:2011ar, Biswas:2012bp, Nojiri:2012zu}.

\subsubsection{Conformal Gravity}

Dropping the first two terms in
Equation \eqref{SHD2}, one obtains the action for conformal gravity
\cite{bach},
\begin{align}
S^{\mathrm{CG}}[g]=-\frac{c_{RR}m_g^{2}}{m^2}\int\dd^4x\sqrt{g}
\left[R^{\mu\nu}R_{\mu\nu}-\frac{1}{3}R^2 \right] \,.
\label{CG}
\end{align}

This is the square of the Weyl tensor (modulo the Euler density) and
is invariant under Weyl scalings of the metric. It can also be
constructed as the gauge theory of the conformal group
\cite{Kaku:1977pa}. The new Weyl invariance removes one of the seven
propagating modes of Equation \eqref{SHD2}, so Equation \eqref{CG} propagates only six
modes \cite{Fradkin:1981iu,Lee:1982cp}. In flat backgrounds, these
appear as a healthy massless spin-2, a ghost-like massless spin-2 and
massless spin-one \cite{Riegert:1984hf}. In de Sitter (dS) backgrounds, the helicity
1 modes and one set of the helicity 2 modes become massive
\cite{Maldacena:2011mk}. Conformal gravity appears in many different
contexts and has played a role in several theoretical developments.
For some recent work, a discussion and brief history, see
\cite{Lu:2011ks, Lu:2013hx, Maldacena:2011mk,Metsaev:2007rw,Mannheim:2011ds,Schmidt:2006jt}.

\subsubsection{Ghost-Free Bimetric Theory}

This is formulated in
terms of two metrics, $\gmn$ and $\fmn$, with an interaction potential
described in Section \ref{BM}. The bimetric action $S[g,f]$ is
ghost-free and propagates seven modes
\cite{Hassan:2011zd,Hassan:2011ea} (\mbox{see also
\cite{Alexandrov:2012yv, Soloviev:2013mia}}). For proportional metrics, $\fmn=c^2\gmn$,
the classical solutions coincide with those of general relativity, and
around such backgrounds, the seven modes split into massless and
massive spin-2 fields \cite{Hassan:2012wr}. Due to the presence of the
massless field, this is not a theory of massive gravity. Rather, it
describes gravity in the presence of a neutral spin-2 field in a
completely dynamical setup. \mbox{The bimetric} spectrum is similar to that
of higher derivative (HD) gravity Equation \eqref{SHD2}, except that neither of the spin-2 fields
is a ghost. For some recent work within bimetric theory, see
\cite{Volkov:2011an,vonStrauss:2011mq, Comelli:2011zm,Berg:2012kn,
 Park:2012cq, Sakakihara:2012iq, Akrami:2012vf,
 Capozziello:2012re, Mohseni:2012ug, Baccetti:2012ge,
 Baccetti:2012re,Baccetti:2012bk,Volkov:2012wp,Myrzakulov:2013owa,
 Maeda:2013bha}.

We refer to the ghost-free bimetric models as the Hassan--Rosen (HR)
models, as distinct from the fixed background de Rham--Gabadadze--Tolley (dRGT)
 models
\cite{deRham:2010ik,deRham:2010kj} (for a recent review of massive gravity, see, e.g.,~\cite{deRham:2014zqa}), to emphasize that they have very
distinct dynamics and physics. The dRGT models, which propagate five
nonlinear modes \cite{Hassan:2011hr,Hassan:2011tf}, are obtainable
from the HR ones by dropping the kinetic term and the equation of
motion for $\fmn$, which is then set to some fixed background metric,
for instance $\fmn=\eta_{\mu\nu}$, as in the original dRGT setup
\cite{deRham:2010ik,deRham:2010kj}. This procedure cannot be
implemented at the level of the bimetric action or the full bimetric
equations of motion around generic backgrounds. It can only be
implemented as a scaling limit on a class of bimetric solutions that
survive in the limit. It is important to realize that when the massive
gravity limit is taken for a certain class of solutions (with properly
scaled parameters, so they remain regular), there also exists a large set
of bimetric solutions that are singular in the massive gravity limit
\cite{Hassan:2014vja}.\footnote{For example, bimetric theory has an
 interesting class of solutions that admit a general relativity limit
 without a vDVZ (van Dam--Veltman--Zakharov)
 discontinuity \cite{Akrami:2015qga}, while no such
 solutions exist in massive gravity.} Therefore, the limit needs to be
treated with care, and statements made in the massive gravity context
do not automatically carry over to the bimetric theory, which has a
richer solution space. In other words, the HR and dRGT setups not
only have different field contents, but could also have very distinct
dynamics and predictions. In particular, this difference is crucial
for understanding that the recent analysis of acausality in massive
gravity \cite{Deser:2012qx, Deser:2014hga} and the arguments for the
absence of partially massless theories in the fixed background setup
\cite{Deser:2013uy,deRham:2013wv} do not automatically apply to the HR
theory. These works rely on the existence of a fixed reference metric
$\fmn$ and, in the present form, manifestly break down in the fully
dynamical bimetric~setup.

\subsubsection{Linear Partially Massless Theory and Beyond}

Partial masslessness was first observed in the Fierz--Pauli theory of a
linear massive spin-2 field in a dS background. In $d=4$, it was
found that when the mass and the cosmological constant satisfy the
Higuchi bound, $m_\mathrm{FP}^2=\frac{2}{3}\Lambda$ \cite{Higuchi:1986py}, the
massive spin-2 field has four propagating modes, instead of the usual
\mbox{five \cite{Deser:2001us}}. The reduction is due to a new gauge symmetry
that emerges at the Higuchi bound. \mbox{The resulting} theory has been
called partially massless. The obvious question is if there exists a
nonlinear PM theory that goes beyond the Fierz--Pauli setup and that
could even give a meaning to \mbox{partial masslessness} away from dS
backgrounds.\footnote{The PM phenomenon also arises in higher spin
 theories~ \cite{Francia:2008hd,Joung:2012rv,Joung:2012hz}, but here,
 we consider only the spin-2 case.}

A perturbative approach to finding such a theory has been to construct
cubic vertices with PM gauge invariance in the linear Fierz--Pauli theory
\cite{Zinoviev:2013hac,Joung:2012hz}. This has led to the insight
that, in a two-derivative theory, cubic PM vertices exist only in three
and four dimensions \cite{Zinoviev:2006im,Joung:2012rv}. In higher
derivative theories, such vertices can also be constructed in higher
dimensions \cite{Joung:2012rv}. However, extending this method beyond
cubic order becomes rather tedious.

Alternatively, a footnote in \cite{Maldacena:2011mk} observed that
conformal gravity Equation \eqref{CG} around dS backgrounds propagates a
massless spin-2 field along with four massive modes that could be
identified with the PM field. One of the two fields is now a ghost,
depending on the overall sign of the action. The possibility of
conformal gravity as a nonlinear PM theory was further investigated in
\cite{Deser:2012qg}. The criterion was to look for a four-component PM
field away from dS or Einstein backgrounds.\footnote{Linear PM modes
 exist in backgrounds specified by $R_{\mu\nu}=\Lambda \gmn$, i.e.,
 in Einstein spacetimes and not just in dS spacetimes. From now on,
 this is assumed to be understood, even when not explicitly stated.}
Such modes were not found, leading to the conclusion that conformal
gravity is not a nonlinear PM theory, though it could still prove
useful to the problem. From the bimetric point of view described below,
the criterion in \cite{Deser:2012qg} is too restrictive, as it requires
that the PM theory around any background looks similar to that around a dS~background.

The main obstacle to a systematic construction of nonlinear PM theory
is the relation to massive spin-2 fields that generically contain a
Boulware--Deser ghost \cite{Boulware:1973my}. Hence, the search must
focus on such ghost-free theories. The first of these is the dRGT
model \cite{deRham:2010ik,deRham:2010kj}, often described as massive
gravity, though it also describes a massive spin-2 field $\gmn$ in a
flat spacetime $\eta_{\mu\nu}$. The model is ghost-free nonlinearly
\cite{Hassan:2011hr}, but it does not admit dS backgrounds
\cite{DeFelice:2013awa} and, hence, does not accommodate even the linear
PM theory. The generalizations of dRGT models to theories of a massive
spin-2 field $\gmn$ in arbitrary curved backgrounds $\fmn$
\cite{Hassan:2011vm} also exist and are free of the BD (Boulware--Deser)
 ghost
\cite{Hassan:2011tf,Hassan:2012qv}. These do admit dS backgrounds for
$\bar g=\lambda^2 f$ and can easily accommodate the known linear PM
theory. A nonlinear PM candidate along these lines had been proposed
in \cite{deRham:2012kf}, but it was subsequently shown that this theory
cannot accommodate an extra nonlinear gauge invariance
\cite{Deser:2013uy,deRham:2013wv}.

\subsubsection{Partially Massless Bimetric Theory}

The HR
bimetric theory \cite{Hassan:2011zd} has dS solutions for
$\fmn=c^2\gmn$, where, in general, $c^2$ is determined in terms of the
parameters $\beta_n$ and $\alpha=m_f/m_g$ of the theory. In such
backgrounds, the expressions for the cosmological constant and the Fierz--Pauli
mass of the spin-2 fluctuations are known
\cite{Hassan:2012wr,Hassan:2012rq}. On imposing the Higuchi bound, one
can easily recover the linear PM theory and its gauge symmetry. Then,
as it turns out, a necessary condition for the consistency of the
linear PM symmetry with the dynamical backgrounds is that the theory
leaves $c^2$ undetermined \cite{Hassan:2012gz,Hassan:2012rq}. This
uniquely determines a specific bimetric theory as the candidate
nonlinear PM theory that exists even away from dS or Einstein
backgrounds.

Although the gauge symmetry away from dS backgrounds is not yet known,
there are indications that it exists, in which case the theory will
propagate six modes (instead of the seven for a generic HR model) even
away from dS backgrounds. However, only around dS backgrounds, these
can be decomposed as four components of the PM field and two
helicities of a massless graviton \cite{Hassan:2012wr}. Otherwise, the six
modes will not have such simple decompositions in terms of mass
eigenstates. Such a behavior is not unexpected from a field theoretical
perspective due to the nonlinear interactions of fields with the same
spin. Hence, our criterion for identifying a nonlinear generalization
of the known PM theory is that it has six propagating modes (instead
of the generic seven), irrespective of how they decompose.\footnote{In
the absence of a nonlinearly identifiable four-component PM multiplet,
one may object to labelling these as PM theories. We use the name at
least for historical reasons, but more importantly, since these are
seemingly the only possible nonlinear generalizations of the linear PM
theories.}

One piece of evidence in favor of the bimetric PM proposal is that, in a two
derivative setup, it correctly predicts nonlinear PM theories only in
$d=3$ and $d=4$. However, on allowing for higher derivative terms, one
obtains candidate PM theories, even in higher dimensions
\cite{Hassan:2012rq}. This is in accordance with the outcome of
the direct construction of PM cubic vertices mentioned above (for some
more arguments, see \cite{Hassan:2012gz,Hassan:2012rq}). Further
evidence for the existence of an extra symmetry in the PM bimetric
theory will be provided below.

It should also be emphasized that, even if it eventually turns out
that the bimetric setup cannot provide a nonlinear description of PM
theory, in the sense that only six degrees of freedom are propagating,
the candidate theory is still worth studying. Its special properties
and, in particular, the connection to conformal gravity that will be
established in this work provide a richer framework for studying
partial masslessness, possibly within an extension of bimetric theory.
It would then be feasible that, in this larger framework, the
additional gauge symmetry does exist, while in bimetric theory,
we only discover remnants of it.

\subsection{Summary of Results} \label{summary}

Our results are summarized below:

{\bf Correspondence between bimetric theory and HD gravity:} Starting
with the bimetric action $S[g,f]$ in terms of spin-2 fields $\gmn$ and
$\fmn$, we determine $\fmn$ algebraically in terms of $\gmn$ and its
curvatures $R_{\mu\nu}(g)$. In general, the solution $\fmn(g)$ is a
perturbative expansion in powers of $R_{\mu\nu}(g)/m^2$, where
$m^2$ sets the scale of the Fierz--Pauli mass in the bimetric theory.
While this expression is valid for $R(g)<m^2$, in special cases, the
series terminates and the solution is exact. Using this solution to
eliminate $\fmn$ from the bimetric action, we obtain the action for
higher derivative gravity as,
\be
S^\mathrm{HD}[g]=S[g,f(g)]\,.
\label{SHD-Sgf}
\ee

At the four-derivative level, this gives $S^\mathrm{HD}_{(2)}$ in
Equation \eqref{SHD2}. We perform this calculation in $d$ dimensions, and hence,
for $d=3$, we can obtain a higher derivative extension of the new
massive gravity setup of \cite{Bergshoeff:2009hq} as dictated by the
bimetric theory.\footnote{For another attempt to obtain new massive
 gravity from bimetric theory, involving certain scaling limits of
 the parameters, see \cite{Paulos:2012xe}. In that approach, one
 introduces spin-2 ghosts already in bimetric theory by taking the
 wrong sign for the $f$-metric kinetic term. In the present paper, we
 keep the bimetric theory healthy and try to pin point the origin of
 the ghost in the higher-derivative theory.}

We also obtain a correspondence between the equations of motion in HD
gravity and bimetric theory. These differ by an extra operator
$(\delta f/\delta g)$. At the linearized level, if $\chi(x)$ solves
the differential equation $(\delta f/\delta g)\chi=0$, then the
solutions of the HD gravity equations have the form
$g^\mathrm{HD}=g^\mathrm{HD}(x,\chi(x))$. \mbox{The classical} solutions in
bimetric theory are related to these by, \be
g^\mathrm{BM}(x)=g^\mathrm{HD}(x, \chi=0)\,. \ee

Thus, on setting
$\chi=0$, say, through boundary conditions, the classical solutions in
the two theories will coincide.

{\bf Truncation and ghost in HD gravity:} The HD gravity action
truncated to quadratic curvature terms, $S^\mathrm{HD}_{(2)}$,
Equation \eqref{SHD2}, has seven propagating modes that, around appropriate
backgrounds, decompose into a massless and a massive spin-2 field.
This is similar to the bimetric spectrum, except for two important
differences. The massive spin-2 field in $S^\mathrm{HD}_{(2)}$ is a
ghost, and its mass differs from the value in the associated bimetric
theory. We illustrate that, in the linear theory, these discrepancies
are the artifacts of truncating HD gravity to a four-derivative
theory. Resolving the ghost issue then also requires using appropriate
source couplings, as dictated by the associated bimetric theory.

To make this explicit, in the Appendix, we consider two examples, the
linearized bimetric theory and a very similar, but much simpler theory
of two scalars fields. In both cases, one of the fields can be
eliminated to explicitly obtain a higher derivative action for the
remaining field, analogous to the untruncated $S^\mathrm{HD}$. This
action contains appropriate modified source couplings inherited from
the bi-field theory. To check for ghosts, we compute the vacuum
persistence amplitude, which turns out to be exactly the same as the
one computed in the associated bi-field theory. Hence, the untruncated
HD theory is ghost free and has the right mass poles. However, one can
see that if this theory is truncated to four-derivative terms, the
massive field becomes a ghost, and its mass shifts away from the
correct value.

{\bf Conformal gravity and the PM bimetric theory:} For the specific
bimetric theory that has been identified as the candidate nonlinear PM
theory \cite{Hassan:2012gz,Hassan:2012rq}, the corresponding HD theory
at the four-derivative level is conformal gravity Equation \eqref{CG}. What is
more, in this case, the correspondence is closer to an equivalence:
while in general, the truncated equations in HD gravity and bimetric
theory are not the same, in the PM case, the lowest order bimetric
equation is identical to the conformal gravity equation, expressed as
the vanishing of the Bach tensor,
\be
B_{\mu\nu}(g)=0\,.
\ee

For the PM bimetric theory, the implication is that, in the limit
$R(g)<< m^2$, it propagates six (rather than seven) modes even away
from dS backgrounds. The new gauge symmetry that exists in the limit
$R(g)<< m^2$, not only around dS backgrounds, but around any Bach flat
background, corresponds to Weyl transformations of $\gmn$. This is a
further indication that the full PM bimetric theory indeed has an
extra gauge symmetry. Since the Bach equation is now derived from the
bimetric action instead of the conformal gravity (CG) action, none of the six modes it
propagates is a ghost.

Conversely, the above equivalence implies that the PM bimetric
theory is a genuine extension of conformal gravity that resolves its
ghost problem. In the process, the Weyl symmetry of CG is replaced by
the PM symmetry of bimetric theory.

From the bimetric perspective, it is obvious that only around
proportional dS backgrounds, the six modes split into a massless
graviton and a four-component PM field. Away from dS backgrounds, the
six modes do not admit such a split \cite{Hassan:2012wr}, and in
particular, it is not possible to freeze gravity and retain only the
PM field, as attempted in \cite{Deser:2012qg}.

\section{Higher Curvature Gravity from Bimetric Theory}

Starting with the bimetric theory, we will now obtain the higher
derivative gravity actions as an expansion in the inverse of the mass
scale. We will explicitly compute quadratic curvature terms after first
outlining the procedure and a brief review of the ghost-free bimetric
theory.

\subsection{Outline of Obtaining Higher Derivative Gravity from
 Bimetric Theory}

The bimetric action $S[g,f]$ involves the two spin-2 fields $\gmn$ and
$\fmn$ with a non-derivative interaction potential to be specified
later. The corresponding equations of motion,
\be
\left.\frac{\delta S}{\delta g^{\mu\nu}}\right|_f=0\,,\qquad
\left.\frac{\delta S}{\delta f^{\mu\nu}}\right|_g=0\,,
\label{eom-gen}
\ee
are coupled differential equations and include the sources
$T_{\mu\nu}^g$ and $T_{\mu\nu}^f$. Eliminating one of the metrics, say
$\fmn$, between the two equations leads to a higher derivative
equation for $\gmn$. Of course, this is just a step in the direction of
solving Equation \eqref{eom-gen}, and the resulting higher derivative equation
is completely within the framework of bimetric theory. However, as will be
exhibited for a scalar field in the Appendix, it is not
straightforward to derive this higher derivative equation directly
from a local ghost-free action for $\gmn$ alone. Therefore, this manipulation
will not lead to standard higher curvature gravity actions.

On the other hand, note that the $\gmn$ equations do not contain
derivatives of $\fmn$. Hence, they can be solved algebraically to
determine $\fmn$ in terms of $\gmn$ and its curvatures, at least
perturbatively. This will yield $f=f(g, T^g)$. Using this to eliminate
$\fmn$ in the bimetric action $S[g,f]$ then leads to a higher
derivative action for $\gmn$, which precisely coincides with the
standard class of higher curvature gravity actions. Before getting
into the details, let us clarify the relation between the
equations of motion in HD gravity and in bimetric theory.

Suppose one obtains $f=f(g)$ by solving either of the equations in
Equation \eqref{eom-gen}. This can be used to convert the bimetric action into
a higher derivative action $S'[g]=S[g,f(g)]$ for $g$. Varying $S'$
with respect to $g$ gives,
\be
\label{effs}
\frac{\delta S'}{\delta g(x)}=\left.\frac{\delta S}{\delta
g(x)}\right|_{f(g)}+\int\,\dd^dy\frac{\delta f(y)}{\delta g(x)}
\left.\frac{\delta S}{\delta f(y)}\right|_g=0\,.
\ee

The first term corresponds to the usual bimetric $g$-equation of
motion Equation \eqref{effs}, while the second term corresponds to the bimetric
$f$ equation of motion, multiplied by a Jacobian. Two cases arise
depending on how $\fmn$ is determined.

If $\fmn$ is a solution to the $f$-equation $\frac{\delta S}{\delta
 f}|_g=0$, say as $f=f(g,T^f)$, then the second term vanishes, and
Equation \eqref{effs} implies exactly the bimetric $g$-equation. However, this
requires solving a differential equation for $\fmn$ (which is not easy
to solve in full generality), and the solution depends on the boundary
conditions. Hence, in this case, the action $S'$ is bi-local, and the
resulting theory is completely equivalent to the bimetric theory. This
approach is not the focus of our attention in the present paper.

Now, suppose that we instead use the bimetric $g$-equation,
$\frac{\delta S}{\delta g}|_f=0$, to algebraically determine
\mbox{$f=f(g,T^g)$}. Since the solution is algebraic, $S'[g]$ is a local
higher derivative action for $g$ and coincides with $S^\mathrm{HD}[g]$,
Equation \eqref{SHD-Sgf}. Later, we will see that, when
expanded in powers of curvatures, it reproduces the standard higher
derivative gravity action Equation (\ref{SHD2d}) at the lowest orders.
From Equation (\ref{effs}), the equation of motion
becomes,
\be\label{S'eom}
\frac{\delta S^\mathrm{HD}}{\delta g(x)}=\int\,\dd^dy
\frac{\delta f(y)}{\delta g(x)} \left.\frac{\delta S}{\delta f(y)}
\right|_g=0\,.
\ee

The Jacobian factor is in fact an operator, $\frac{\delta f(y)}
{\delta g(x)}=\delta(x-y)\hat{\cal O}_y$, so the higher derivative
$g$-equation of motion is,
\be
\hat{\cal O}\left(\frac{\delta S[g,f]}{\delta f}\Big|_g\right)
=0\,,
\ee
where $f=f(g,T^g)$. This differs form the corresponding bimetric
equation by the presence of the extra operator $\hat{\cal O}$. For
functions $\chi$, such that $\hat{\cal O}\chi=0$, we get,
\be
\frac{\delta S[g,f]}{\delta f}\Big|_g=\chi(x)\,,
\ee

Thus, from the bimetric point of view, $\chi(x)$ appears as an unusual
source term specified through boundary conditions. Let us denote a
solution of this equation by $g^\mathrm{HD}(x, \chi(x))$.
Thus, only for boundary conditions that give $\chi=0$, the solutions of
the higher curvature action $S^\mathrm{HD}[g]$ coincide with the
solutions of bimetric theory, $g^\mathrm{BM}(x)=g^\mathrm{HD}(x,
\chi=0)$.

The difference between the bimetric and HD gravity equations,
contained in $\hat{\cal O}$ is easy to isolate formally. However, in
the truncated theory, $\hat{\cal O}$ mixes with $\delta S/\delta f$,
and generically, the truncated equations are not equivalent (an
exception being conformal gravity). In particular, the higher
derivative theory, when truncated to four-derivative terms, will
contain spin-2 ghosts. However, as the scalar field example in the Appendix
illustrates, these are artifacts of the truncation and of using a
naive source coupling instead of the one dictated by the parent
bimetric theory. Since the truncation scale is given by the mass
scale, the truncated theory is meaningful only at energies below the
mass scale. Conversely, the bimetric theory seems to provide a
ghost-free completion of quadratic curvature gravity.

\subsection{Review of Ghost-Free Bimetric Gravity} \label{BM}

Let us briefly review the relevant equations of HR bimetric theory
\cite{Hassan:2011zd}. More details can be found \mbox{in
\cite{Hassan:2011vm,Hassan:2012wr}} and in the appendix of
\cite{Hassan:2012rq}. The bimetric action is,
\be
S[g,f]=m_g^{d-2}\int\dd^dx\left(\sqrt{g}~R(g)+\alpha^{d-2}\sqrt{f}~R(f)
-2m^2\sum_{n=0}^d\beta_ne_n(S) \right)\,,
\label{bimd}
\ee
where $S$ stands for the square-root matrix ${S^\mu}_\nu=
{(\sqrt{g^{-1}f})^\mu}_\nu$ and $e_n(S)$ are the elementary symmetric
polynomials of the eigenvalues of $S$. Starting with $e_0(S)=1$, they
are constructed iteratively through,
\be
e_n(S)=-\frac{1}{n}\sum_{k=1}^n (-1)^k \Tr(S^k)\, e_{n-k}(S)\,.
\ee

In $d$ dimensions, $e_d(S)=\det{S}$ and $e_n(S)=0$ for $n>d$. We will
explicitly need,
\be
e_1(S)=\Tr S\,,\qquad
e_2(S)=\frac{1}{2}\left[(\Tr S)^2- (\Tr S^2)\right]\,.
\label{e12}
\ee

In Equation \eqref{bimd}, $\alpha=m_f/m_g$ is the ratio of the Planck masses,
and $\beta_n$ are $d+1$ dimensionless free parameters. $m^2$, which is
degenerate with the overall scale of the $\beta_n$, sets the mass
scale of the massive mode. The action retains its form under the
interchange of $g$ and $f$ due to the properties of the $e_n(S)$
\cite{Hassan:2011zd}. The equations of motion for $\gmn$ and $\fmn$
are,
\begin{align}
&\Rmn(g)-\frac{1}{2}\delta^\mu_\nu R(g) +
m^2\sum_{n=0}^{d-1}(-1)^n\beta_n \mathbb{Y}_{(n)\nu}^{\mu}(S)
=m_g^{2-d} {T^g}^\mu_{~\nu}\,,
\label{bmeom-g}\\
&\Rmn(f)-\frac{1}{2}\delta^\mu_\nu R(f) +
\frac{m^2}{\alpha^{d-2}}\sum_{n=1}^{d}(-1)^n\beta_n
\mathbb{Y}_{(n)\nu}^{\mu}(S^{-1}) =m_f^{2-d} {T^f}^\mu_{~\nu}\,,
\label{bmeom-f}
\end{align}
where the matrices $\mathbb{Y}_{(n)}$ are given by (see, for example,
\cite{Hassan:2012rq}),
\beqn\label{ydef}
\mathbb{Y}_{(n)}(S)=\sum_{k=0}^n(-1)^ke_k(S)\,S^{n-k}\,.
\eeqn

Around proportional backgrounds, $\fmn=c^2\gmn$, the degrees of freedom
propagated by the action Equation \eqref{bimd} split into a massless and a
massive spin-2 field. In the present conventions, the Fierz--Pauli mass
is given by \cite{Hassan:2012wr},
\be
m^2_\mathrm{FP}=m^2\left(\frac{1+(\alpha c)^{d-2}}{(\alpha c)^{d-2}}\right)
\sum_{n=1}^{d-1}{d-2\choose n-1} c^n\,\beta_n \,.
\label{FP}
\ee
In general, $c$ is determined in terms of $\beta_n$ and $\alpha$
through the equality of the cosmological constants,
$\Lambda_g=\Lambda_f$. Furthermore, if $\Lambda_g=0$, then Equation \eqref{FP}
gives the mass around flat spacetime.

\subsection{The Algebraic Solutions for $S$ and $f$}

The $g$-equation, Equation \eqref{bmeom-g}, is algebraic in $S^\mu_{~\nu}$ and,
hence, in $\fmn$, so it can be used to determine $\fmn$ in terms of
$\gmn$ its curvatures and $T_{\mu\nu}^g$. First, we consider the minimal
$\beta_1$ model (with all $\beta_n$ for $n\geq2$ vanishing) where the
solution can be found exactly and then find a perturbative solution in
powers of the curvatures for general $\beta_n$.

\subsubsection{Exact Solution in the $\beta_1$ Model:}

In the
$\beta_1$ model (where $\beta_2=\cdots= \beta_{d-1}=0$), the $\gmn$
equation of motion becomes,
\be
\Gmn(g)+
m^2\left[\beta_0\,\delta^\mu_\nu-\beta_1\left(S^\mu_{~\nu}-\delta^\mu_\nu
\Tr S\right)\right]
=m_g^{2-d} {T^g}^\mu_{~\nu}\,,
\ee
where $G_{\mu\nu}=\rmn-\frac{1}{2}\gmn R$ is the Einstein tensor.
The trace of this equation determines $\Tr S$. Then, for
${T^g}^\mu_{~\nu}=0$, one finds the solution,
\be
S^\mu_{~\nu}=-\frac{1}{d-1}\frac{\beta_0}{\beta_1}\,\delta^\mu_\nu
+\frac{1}{m^2\beta_1}\,\Pmn\,,
\label{Ssol-beta1}
\ee

Here, $\Pmn=g^{\mu\lambda}P_{\lambda\nu}$ is the Schouten tensor of
$\gmn$ in $d$ dimensions,\footnote{The conventionally normalized
 Schouten tensor is $\frac{1}{d-2}P_{\mu\nu}$, but the
 non-standard normalization of Equation \eqref{Pdef} is more convenient for
 our purposes}
\beqn\label{Pdef}
\pmn=\rmn-\frac{1}{2(d-1)}\gmn R\,.
\eeqn

From $S$, one can easily find $\fmn=g_{\mu\lambda}
(S^2)^\lambda_{~\nu}$. The solution has the form of a power series in
$P/m^2$ that terminates at the first order.

The solution for $S$ in the presence of matter sources is easily
obtained from Equation \eqref{Ssol-beta1} by the replacement
$G_{\mu\nu}\rightarrow
G_{\mu\nu}-m_g^{2-d}\,T^g_{\mu\nu}$, or equivalently by,
\be
\Pmn\rightarrow \Pmn - m_g^{2-d}\left({T^g}^\mu_{~\nu}-\frac{1}{d-1}
{T^g}^\lambda_{~\lambda} \delta^\mu_\nu) \right)\,.
\label{addT}
\ee

\subsubsection{Perturbative Solution for General $\beta_n$:}

For general $\beta_n$ in Equation \eqref{bmeom-g}, finding an exact solution
for $S=\sqrt{g^{-1}f}$ is not feasible. However, one can always find a
perturbative solution as an expansion in $m^{-2}$. In the
absence of matter sources, this corresponds to a derivative expansion
in powers of $R_{\mu\nu}(g)/m^2$, or equivalently, in
$P_{\mu\nu}(g)/m^2$. In order to find the perturbative
solution, we make the general ansatz,\footnote{Before proceeding further, note that this ansatz captures the generic
class of solutions where a diagonal $R^\mu_{~\nu}(g)$ implies a
diagonal $S^\mu_{~\nu}$. One may also find special bimetric solutions
where a diagonal $R^\mu_{~\nu}(g)$ corresponds to a non-diagonal
$S^\mu_{~\nu}$. To see this, note that Equation \eqref{bmeom-g} has
the generic form $R^\mu_{~\nu}=\sum_r q_r (S^r)^\mu_{~\nu}$, where the
scalars $q_r$ are functions of the $\beta_n$ and of $e_n(S)$. It is
possible to construct non-diagonal $S$, such that $\sum_r q_r S^r$ is
diagonal. The off-diagonal structure depends on the choice of the
$\beta_n$, and this is not possible in models with only one out of
$\beta_1$, $\beta_2$ and $\beta_3$
nonvanishing~\cite{Hassan:2014vja}. For solutions of this type, see,
for example, \cite{Volkov:2012zb}. \label{non-diag}}
\begin{align}
\label{ansatzs}
\Smn=a\delta^\mu_\nu &+\frac{1}{m^2}\left(b_1\Pmn+b_2\Tr
P\,\delta^\mu_\nu \right)\nn\\
&+\frac{1}{m^4}\left(c_1{\Pmn}^2+c_2\Pmn\Tr P+c_3\Tr (P^2)\delta^\mu_\nu
+c_4(\Tr P)^2\delta^\mu_\nu\right)~+~\mathcal{O}(m^{-6})\,.
\end{align}
The coefficients $a, b_i$ and $c_i$ are determined by plugging this
into Equation \eqref{bmeom-g} and matching the coefficients of
different powers of $\Pmn$. The details are given in the Appendix. The
result can be presented simply in terms of the sums, \be\label{sums}
s_k\equiv\sum_{n=k}^{d-1}{d-1-k\choose n-k}\beta_na^n\,, \qquad
k=0,1,2,3\hdots \ee

The coefficient $a$ is determined in terms of
$\beta_n$ as a solution of the polynomial equation, \be\label{eq1or}
s_0\equiv\sum_{n=0}^{d-1}{d-1 \choose n}\beta_n a^n=0\,. \ee

Non-real
solutions are not excluded as long as the parameters in the final
action are real. For the $b_i$, one obtains, \be\label{eq2or}
b_2=0\,,\qquad b_1=\frac{a}{s_1}\,. \ee where $s_1$ is defined in
Equation (\ref{sums}). Finally, the $c_i$ are given by, \be
c_1=-c_2=\frac{as_2}{s^3_1}\,,\qquad c_3=-c_4
=-\frac{as_2}{2(d-1)s^3_1} \,. \ee

These coefficients exist only if
$s_1\neq 0$ (the case $s_1=0$ requires separate treatment and implies
Ricci flat $\gmn$, \emph{cf.}~Equation \eqref{app: Pfirst}), and in terms of them, the
solution for $\Smn$ to $\mathcal{O}(m^{-4})$ reads, \be\label{ssol}
\Smn=a\delta^\mu_\nu+\frac{a}{s_1m^2}\Pmn
+\frac{as_2}{s^3_1m^4}\left[\Big({\Pmn}^2-P\Pmn\Big)+\frac{1}{d-1}e_2(P)
 \delta^\mu_\nu\right]+\mathcal{O}(m^{-6})\,. \ee

Note that the
higher order curvature terms are suppressed by the mass scale, \beqn
m^2s_1=m^2\sum_{n=1}^{d-1}{d-2 \choose n-1}\beta_n a^n\,. \eeqn

This
is proportional to the Fierz--Pauli mass Equation \eqref{FP} of the massive
spin-2 fluctuation on proportional backgrounds $\fmn=a^2\gmn$ and with
vanishing cosmological constant.\footnote{For backgrounds given by
 $\fmn=a^2\gmn$, the cosmological constant $\Lambda_g$ in the
 $g$-equation is given by $s_0$. In the present setup, $a$ must be
 chosen, such that $s_0=0$. For backgrounds $\fmn=c^2\gmn$, with
 $c\neq a$, the cosmological constant is no longer given by $s_0$.}
Thus, the perturbative expansion becomes more accurate for large
Fierz--Pauli mass, although the procedure does not really correspond to
integrating out the massive mode of the bimetric theory. Rather, we
are eliminating $\fmn$, which is not a mass eigenstate.

This procedure can be continued to compute the solution for $\Smn$ to
any order in $m^{-2}$. Here, we are mostly interested in the quadratic
curvature terms in the higher-derivative action for $\gmn$, and hence,
it suffices to determine the solution for $\Smn$ to second order.
As in the $\beta_1$ model, matter sources can be taken into account
through the replacement Equation \eqref{addT} in the final solution. It is easy
to verify that for $\beta_2=\cdots= \beta_{d-1}=0$, one reproduces the
exact result for the $\beta_1$ model obtained above as in this case,
$s_n=0$ for $n\geq 2$.

\subsection{Higher Derivative Gravity from Bimetric Theory}

In this section, we eliminate $\fmn$ in the bimetric action using
Equation (\ref{ssol}) to obtain a higher curvature theory for $\gmn$. We
explicitly retain only quadratic curvature terms and neglect the
source terms, which can always be reinstated by the replacement
Equation \eqref{addT}.

Let us begin by eliminating $S$ from the bimetric interaction
potential. For this purpose, it is convenient to first express the
potential in terms of,
\be
M^\mu_{~\nu}\equiv \frac{1}{a}\Smn-\delta^\mu_\nu\,.
\ee

Then, using the properties of the $e_n(S)$, one has,
\be
2m^2\sqrt{g}\,\sum_{n=0}^{d}\beta_ne_n(S)=2m^2\sqrt{g}\,
\sum_{n=0}^d\alpha_ne_n\left(M\right)\,,
\ee
where we have defined,
\be
\alpha_n=\,\sum_{k=n}^{d}{d-n \choose k-n}a^{k}\beta_k\,.
\label{alphan}
\ee

Since $M$ directly starts at order $m^{-2}$, only $e_n(M)$ with $n\leq
2$ contribute to order $m^{-2}$ in the potential. Explicitly, $e_0(M)=1$
while $e_1(M)$ and $e_2(M)$ are given by,
\begin{align}
e_1(M)&=\frac{1}{s_1 m^2}\,e_1(P) - \frac{s_2}{s_1^3 m^4}
\frac{d-2}{d-1} \,e_2(P) + \mathcal{O}(m^{-6})\,,\nn\\
e_2(M)&= \frac{1}{s_1^2 m^4}\, e_2(P)+ \mathcal{O}(m^{-6})\,.
\end{align}

Hence, the potential is given by,
\begin{align}
&2m^2\sqrt{g}\,\sum_{n=0}^{d}\beta_ne_n(S)= \nn\\
&\quad\qquad 2m^2\sqrt{g}\left(\alpha_0+\frac{\alpha_1}{s_1m^2}\,e_1(P)
+ \frac{1}{s_1^3 m^4}\left(s_1\alpha_2-s_2\alpha_1\frac{d-2}{d-1}
\right)\,e_2(P)\right)+\mathcal{O}(m^{-4})\,.
\end{align}

Using Equation \eqref{Pdef}, it is straightforward to express $e_n(P)$ in terms
of curvatures of $\gmn$,
\be
e_1( P)=\frac{d-2}{2(d-1)}R\,,\qquad e_2( P)=\frac{1}{2}
\left(\frac{d}{4(d-1)}R^2-R^{\mu\nu}R_{\mu\nu}\right)\,.
\ee

Then, the bimetric potential expressed in terms of curvatures becomes,
\begin{align}\label{poteff}
2m^2\sqrt{g}\,\sum_{n=0}^{d}\beta_ne_n(S)
=2m^2 &\sqrt{g}\Bigg[\alpha_0+\alpha_1\frac{d-2}{d-1}\frac{1}{2s_1m^2}R
\nn\\
&+\frac{s_1\alpha_2-s_2\alpha_1\frac{d-2}{d-1}}{2s_1^3m^4}
\Big(\frac{d}{4(d-1)}R^2-R^{\mu\nu}R_{\mu\nu}\Big)\Bigg]
+\mathcal{O}(m^{-4})\,.
\end{align}

Let us now consider the kinetic term $\sqrt{f}R(f)$. In contrast with
the bimetric potential, which produces polynomials only in curvatures of
$\gmn$, the kinetic term also produces terms with covariant
derivatives acting on curvatures. At quartic order in
derivatives, the Bianchi identity can be used to show that the only
possible such term is $\nabla^2 R$. However, it will turn out in the
following that the coefficient of this term is zero for all
$\beta_n$. Terms involving derivatives of curvatures only start to
appear at cubic order.

The solution for $\fmn$ is obtained from Equation (\ref{ssol}) using
$\fmn=g_{\mu\rho}{(S^2)^\rho}_\nu$ and reads, to first order \mbox{in curvatures},
\be
\fmn=a^2\gmn+\frac{2a^2}{s_1m^2}\pmn
+\mathcal{O}(m^{-4})\,.
\label{f-sol}
\ee

Note that in $\sqrt{f}R(f)$, only terms up to this order contribute to
the quadratic curvature terms. \mbox{Expanded to} first order in powers of
$m^{-2}$, the inverse of $\fmn$ is then given by,
\be
(f^{-1})^{\mu\nu}=a^{-2}g^{\mu\nu}-\frac{2}{a^2s_1m^2}P^{\mu\nu}
+\mathcal{O}(m^{-4})\,.
\ee

Using the curvature relations (see, e.g.,~\cite{Wald:1984rg}),
\begin{align}
R_{\mu\nu}(f)&=R_{\mu\nu}(g)+2\nabla_{[\mu}{C_{\alpha]\nu}}^\alpha-
2{C_{\nu[\mu}}^\beta{C_{\alpha]\beta}}^\alpha\,, \nn\\
{C_{\mu\nu}}^\alpha&=\tfrac{1}{2}(f^{-1})^{\alpha\rho}\left(\nabla_\mu
f_{\nu\rho}+ \nabla_\nu f_{\mu\rho}-\nabla_\rho f_{\mu\nu}\right) \,,
\end{align}
we can express the curvatures of $\fmn$ in terms of curvatures and
connections of $\gmn$. For the $\fmn$ obtained above, this gives,
\be
R_{\mu\nu}(f)=R_{\mu\nu}(g)+\frac{2}{s_1m^2}\left(\nabla_\mu\nabla_\nu
{P^\alpha}_\alpha-\nabla^\rho\nabla_\nu
P_{\rho\mu}-\nabla^\rho\nabla_\mu P_{\rho\nu} +\nabla^2 \pmn \right)
+\mathcal{O}(m^{-4})\,.
\label{Rmnf}
\ee

The ${C_{\mu\nu}}^\alpha$ contribute the terms linear in $\pmn$, and
in $R(f)=f^{\mu\nu}R_{\mu\nu}(f)$, these drop out due to the usual Bianchi
identity,
\begin{align}
R(f)&=a^{-2}R(g)-\frac{2}{a^2s_1m^2}P^{\mu\nu}R_{\mu\nu}(g)+\frac{4}{a^2s_1m^2}
\left( \nabla^2{P^\alpha}_\alpha-\nabla^\mu\nabla^\nu\pmn\right)
+\mathcal{O}(m^{-4})\nn\\
&=a^{-2}R(g)-\frac{2}{a^2s_1m^2}\left(R^{\mu\nu}R_{\mu\nu}-\frac{1}{2d-2}
R^2\right)+\mathcal{O}(m^{-4})\,,
\label{Rf}
\end{align}
where we have used the identity $\nabla^2{P^\alpha}_\alpha-\nabla^\mu
\nabla^\nu\pmn=0$, which follows from the Bianchi identity. \mbox{As
promised}, the terms involving derivatives acting on curvatures have
dropped out. Furthermore, using \mbox{$\det(1+A)=\sum_n e_n(A)$} and retaining the
first two terms, we have,
\be
\sqrt{f}=a^d\sqrt{g}\,\sum_{n=0}^{d}e_n\left(\frac{P}{s_1m^2}\right)
=a^d\sqrt{g}\left(1+\frac{1}{2s_1m^2}\frac{d-2}{d-1}R(g)\right)
+\mathcal{O}(m^{-4})\,.
\ee

Then, the kinetic terms become,
\begin{align}\label{kineff}
&\sqrt{g}~R(g)+\alpha^{d-2}\sqrt{f}~R(f)\nn\\
&\quad\qquad =\sqrt{g}\left[ \left(1+(\alpha a)^{d-2}\right)R(g)
-\frac{2(\alpha a)^{d-2}}{s_1m^2}\left(R^{\mu\nu}R_{\mu\nu}
-\frac{d}{4(d-1)}R^2\right) \right]+\mathcal{O}(m^{-4})\,.
\end{align}

Note that this has the same structure as the potential terms. Finally,
combining Equations (\ref{poteff}) and~(\ref{kineff}), we obtain the higher
derivative Lagrangian for $\gmn$ to quadratic order in curvatures,
\be
{\mathcal L}^{\mathrm{HD}}_{(2)}(g)
=m_g^{d-2}\sqrt{g}\Bigg[\Lambda +c_R R(g) -\frac{c_{RR}}{m^2}
\left(R^{\mu\nu}R_{\mu\nu}-\frac{d}{4(d-1)}R^2\right)\Bigg]
+\mathcal{O}(m^{-4})\,.
\label{LHD2}
\ee

This coincides with Equation \eqref{SHD2}, and the parameters are given by,
\begin{align}
\Lambda&=-2m^2\alpha_0 \,,\nn\\
c_R&=1+(\alpha a)^{d-2}-\frac{\alpha_1}{s_1}\,\frac{d-2}{d-1}\,,
\nn\\
c_{RR}&= \frac{1}{s_1^2}\left(2(\alpha a)^{d-2}s_1-\alpha_2
+\frac{s_2}{s_1} \frac{d-2}{d-1}\alpha_1\right)\,.
\label{cs}
\end{align}

The theory can also be written in a more compact form,
\be
{\mathcal L}^{\mathrm{HD}}_{(2)}(g)
=m_g^{d-2}\sqrt{g}\Bigg[\Lambda+ c'_R\,e_1( P)
+c'_{RR}\, e_2(P)\Bigg] +\mathcal{O}(m^{-4})\,,
\label{LHD2e}
\ee
where $c'_R=2(d-1)c_R/d-2$ and $c'_{RR}=2c_{RR}/m^2$.

An identity that
follows from Equations \eqref{sums} and \eqref{alphan} is,
\be
s_n=\alpha_n-\alpha_{n+1}\,.
\label{saa}
\ee

Since $s_0=0$, this implies $\alpha_0=\alpha_1$ and also gives an
expression for the coefficients in terms of the $\alpha_n$ or using
Equation \eqref{alphan} in terms of the $\beta_n$.
Note that ${\mathcal L}^{\mathrm{HD}}_{(2)}$ depends on only three
combinations of the $d+1$ parameters $\beta_n$. To see all of the free
parameters of the bimetric theory, one needs to consider higher order
curvature terms in ${\mathcal L}^{\mathrm{HD}}$. The truncation is
accurate only in the low curvature regime $R(g)<< m^2$. It is worth
pointing out, however, that although the HD Lagrangian is in general an
infinite expansion in curvatures, the number of parameters determining
this expansion is just the $d+1$ parameters $\beta_n$ together with
$\alpha=m_f/m_g$.

\section{The Ghost Issue and Relevance to New Massive Gravity}

The higher derivative gravity Lagrangian \eqref{LHD2} propagates seven
modes in $d=4$, comprising two modes of a healthy massless graviton and
five modes of a massive spin-2 field, which is a ghost. This is in contrast
to the original bimetric theory in which all seven modes are healthy. The
masses of the spin-2 fields are also not equal in the two theories. Of
course, in the HD gravity case, one is working with a truncated action
where the mass of the problematic excitation is related to the
truncation scale. Therefore, it is expected that the ghost problem is an
artifact of the truncation.

To illustrate this, in the Appendix, we consider two examples, the
linearized version of bimetric theory and a very similar, but much
simpler theory of two scalars fields. In the absence of interactions,
these theories can be manipulated explicitly. In both the spin-0 and
spin-2 examples, one of the fields can be eliminated to explicitly
obtain a closed form higher derivative action for the remaining field,
including appropriate source couplings inherited from the original
bi-field theory. The action contains up to six-derivative terms. To check
for ghosts, we compute the vacuum persistence amplitude in the higher
derivative theory. This turns out to be exactly the same as the one
computed in the associated bi-field theory. The equivalence implies
that the untruncated HD theory is ghost free and has the right mass
poles. However, if this theory is now truncated to four-derivative
terms, the massive field becomes a ghost, and its mass shifts away from
the correct value, thus illustrating the point.

In the presence of bimetric interactions, the corresponding HD gravity
obtained in a perturbative manner will not have a finite number of terms,
and it is not feasible to write such a theory in a closed form.
Considering this, $S^\mathrm{HD}_{(2)}$, Equation \eqref{SHD2}, is usable only as
an effective theory at energy scales below the mass pole, and the
$R^2$ terms must be treated only perturbatively. The present analysis,
however, implies that $S^{\mathrm{HD}}_{(2)}$ has a minimal ghost-free
completion in the form of an HR bimetric theory. Even simpler, since
at the four-derivative level, the HD action has only three independent
parameters, its structure can be reproduced by the $\beta_1$ model
alone. This will also avoid the issue raised in footnote \ref{non-diag}.
\mbox{The equations} of motion derived from the full
$S^\mathrm{HD}$ differ from the corresponding bimetric equation (obtained
on eliminating $\fmn$ between Equations~ \eqref{bmeom-g} and \eqref{bmeom-f}) by
the extra operator $\delta f/\delta g$, as discussed earlier. The
truncation procedure mixes this operator with the HD equation of
motion, and the similarity with the bimetric equation is no longer
obvious.

The correspondence between the HR bimetric theory and the action
$S^\mathrm{HD}$ may also be regarded as a reconfirmation that the
bimetric theory propagates only seven modes and no Boulware--Deser
ghost \cite{Boulware:1973my}.

One can also consider the implications of this correspondence to new
massive gravity (NMG). \mbox{There has} been some work to discover and
understand ghost-free higher dimensional generalizations of the NMG
models, with no success. NMG corresponds to ${\mathcal
 L}^{\mathrm{HD}}_{(2)}$ for $d=3$ and $m_g\,c_R<0$, in which case the
theory contains a healthy massive spin-2 mode. With $m_g\,c_R<0$, the
massless spin-2 mode becomes a ghost, but in $d=3$, this mode does not
propagate. This way, one ends up with a healthy theory of a massive
spin-2 field. However, this mechanism of removing the ghost is
peculiar to $d=3$, and attempts to generalize NMG beyond $d=3$ have so
far remained largely unsuccessful \cite{Kleinschmidt:2012rs}. The
present analysis indicates that the bimetric theory is indeed one
(possibly the only) minimal completion of NMG, providing a ghost-free
generalization to all dimensions.

\section{Conformal Gravity from Partially Massless Bimetric Theory}

Recently, the problem of finding nonlinear realizations of partial
masslessness, first observed in the linear Fierz--Pauli theory in dS
backgrounds \cite{Deser:2001us}, has attracted some attention
\cite{Zinoviev:2006im,Joung:2012rv,deRham:2012kf,Deser:2012qg,
 Hassan:2012gz,Hassan:2012rq}. \mbox{One such} nonlinear candidate is the PM
bimetric theory that exists only in three and four dimensions and that
directly contains the original linear PM theory in dS backgrounds
\cite{Hassan:2012gz,Hassan:2012rq}. In these works, it was not
established that away from dS backgrounds, this theory still contains
six modes (in $d=4$), rather than the generic seven modes of the HR
bimetric theory. Another possible PM candidate is conformal gravity,
whose spectrum around dS backgrounds was observed to be similar to the
linear PM spectrum~\cite{Maldacena:2011mk}. \mbox{CG in} general propagates
six modes, but the spectrum is plagued by the usual ghosts of HD
gravity. \mbox{On the} face of it, these are very different approaches to the
PM issue.

It is now easy to check that the HD gravity corresponding to the PM
bimetric theory is indeed related to conformal gravity. What is more,
it turns out that at the four-derivative level, the PM bimetric theory
and CG have exactly the same equations of motion (which is not the
case for the non-PM actions considered above). Hence, the PM bimetric
theory is indeed a ghost-free extension of conformal gravity.
Conversely, this provides further evidence that the PM bimetric theory
propagates six modes even away from \mbox{dS backgrounds}.

\subsection{The Correspondence in $d=4$}\label{actionlevel}

The partially massless bimetric theory in $d=4$ is specified by
setting \cite{Hassan:2012gz,Hassan:2012rq},
\be\label{pmpar}
\beta_1=\beta_3=0\,,\qquad \alpha^4\beta_0=3\alpha^2\beta_2=\beta_4\,.
\ee

Then, Equations \eqref{sums}, \eqref{eq1or} and \eqref{alphan} give,
\be
a^2=-\frac{1}{\alpha^2}\,, \qquad s_1=-\frac{2\beta_2 }{\alpha^2}\,,
\qquad \alpha_0=\alpha_1=0\,,\qquad \alpha_2=\frac{2\beta_2}{\alpha^2}\,.
\ee
and the solution for $\fmn$ takes the form,
\be
\fmn=-\frac{1}{\alpha^2}\gmn+\frac{1}{\beta_2 m^2}\pmn
+\mathcal{O}(m^{-4})\,.
\label{pmf-sol}
\ee

Using these parameter values in Equation \eqref{cs}, the higher derivative
Lagrangian Equation \eqref{LHD2} reduces to,
\begin{align}\label{cgact}
\mathcal{L}^\mathrm{CG}
=-\frac{\alpha^2m_g^2}{2\beta_2m^2}\sqrt{g}
\left(R^{\mu\nu}R_{\mu\nu}-\frac{1}{3}R^2\right) +\mathcal{O}(m^{-4})\,.
\end{align}

Restricted to quadratic order in curvatures, this is the well-known
action for conformal gravity. It is invariant under Weyl scalings of
the metric, and as a result, it contains only four propagating massive
modes, in addition to two massless modes.

The relation between conformal gravity and PM bimetric theory is
constrained enough that the above reasoning can be carried out in
reverse: demanding that the higher derivative action Equation \eqref{LHD2}
reduces to conformal gravity at the quadratic level (that is,
$\alpha_0=0$, $c_1=0$) fixes the parameters of the bimetric theory to
their PM values. Explicitly, $\alpha_0=0$ and $s_0=0$ Equation (\ref{eq1or})
imply $\alpha_1=0$, since $\alpha_0-\alpha_{1}=s_0$, Equation (\ref{saa}).
Hence, in order for the Einstein--Hilbert term to vanish, that is
$c_1=0$, we must have $a^2=-\alpha^{-2}$. Then, the $s_0=0$ equation,
Equation (\ref{eq1or}), becomes,
\beqn
\beta_0+\frac{3i}{\alpha}\beta_1-\frac{3}{\alpha^2}\beta_2-
\frac{i}{\alpha^3}\beta_3=0\,.
\eeqn

Furthermore, the condition $\alpha_0=0$ reads, Equation \eqref{alphan},
\beqn
\beta_0+\frac{4i}{\alpha}\beta_1-\frac{6}{\alpha^2}\beta_2
-\frac{4i}{\alpha^3}\beta_3+\frac{1}{\alpha^4}\beta_4=0\,.
\eeqn

The real and imaginary parts must vanish simultaneously, which gives
Equation (\ref{pmpar}) as the unique solution. This establishes the
correspondence between CG and PM bimetric theory.

\subsection{A Step Further: Equivalence between CG and PM Bimetric
 Theory} \label{eomlevel}

As we have seen, in general, the bimetric equations of motion and the
corresponding HD gravity equations differ by an operator $\delta
f/\delta g$. On truncating to four-derivative terms, as
in $S^\mathrm{HD}_{(2)}$, one finds that the truncated theories are
not equivalent. An exception is the PM case, as shown below.

The equations of motion arising from the conformal gravity action
Equation (\ref{CG}) correspond to setting the Bach tensor $B_{\mu\nu}$ of
$\gmn$ to zero \cite{bach},
\be
B_{\mu\nu}\equiv -\nabla^2
P_{\mu\nu}+\nabla^\rho\nabla_{(\mu}P_{\nu)\rho}+
W_{\rho\mu\nu\sigma}P^{\rho\sigma}=0\,.
\label{CGeom}
\ee

In this expression, the four-dimensional Weyl tensor
$W_{\rho\mu\nu\sigma}$ is given by:
\be
W_{\rho\mu\nu\sigma}\equiv
R_{\rho\mu\nu\sigma}+g_{\mu[\nu}R_{\sigma]\rho}-g_{\rho[\nu}R_{\sigma]\mu}
+\frac{1}{3}g_{\rho[\nu}g_{\sigma]\mu}R\,.
\label{back}
\ee

Let us now consider the corresponding equation in the PM bimetric
theory. The bimetric $g$-equation has already been solved to
determine $S$ Equation \eqref{ssol} and $\fmn$ Equation \eqref{f-sol}. We substitute
these in the bimetric $f$-equation, Equation \eqref{bmeom-f}, using the
expression Equations \eqref{Rmnf} and \eqref{Rf} for the curvatures of
$\fmn$. This gives a higher derivative equation for $\gmn$. On
restricting to the PM parameter values, \mbox{Equation \eqref{pmpar}}, it is
straightforward to show that the cosmological constant term and the
Einstein tensor drop out, and the equation reduces to,
\be
-B_{\mu\nu} + \mathcal{O}(m^{-4})=0\,.
\label{bm-hdeom}
\ee

Hence, the PM bimetric equation, evaluated to lowest non-trivial
order in $R_{\mu\nu}(g)/m^2$, coincides with the CG
equation.\footnote{This is, of course, consistent with the general
 analysis of the previous sections and follows from the fact that, in
 the PM case, only the lowest order term in $\delta f/\delta g\sim
a^2=-1/\alpha^2$ contributes to the relations between the equations of
motion.}

In other words, the conformal gravity Equation \eqref{CGeom} is the
genuine low curvature limit of the PM bimetric theory. The difference
is that while some of the propagating modes arising from this equation
have negative norms in the CG framework, they appear as healthy
non-ghost modes in the HR bimetric setup. This implies that the PM
bimetric theory can be regarded as a ghost-free extension of \mbox{conformal
gravity}.

This equivalence also has implications for the PM bimetric theory that
has been identified as the only possible nonlinear generalization of
the linear PM theory in the HR bimetric setup \cite{Hassan:2012gz}. It is
known that this theory has six propagating modes around dS
backgrounds, a massless spin-2 field and the four-component PM field.
The seventh mode, which exists in a generic HR bimetric theory, is
eliminated by a new gauge invariance \cite{Deser:2001us}. The
question is if away from dS backgrounds, this theory will still
have a gauge invariance and, consequently, propagate only six
modes. Some indirect evidence for this has been discussed in
\cite{Hassan:2012gz,Hassan:2012rq}. The relation to conformal gravity
found here provides direct evidence for the proposal, as it establishes
that, in the small curvature limit, the PM bimetric theory indeed
propagates only six modes {around any background}.

The Bach Equation (\ref{CGeom}) propagates six modes not only around
dS \cite{Maldacena:2011mk}, but also around a flat \mbox{background
\cite{Riegert:1984hf}}. Thus, an obvious question is whether the higher
derivative corrections to the Bach equation in Equation (\ref{bm-hdeom}) can
alter the propagator for the linear fluctuation around a flat
background. \mbox{We will} argue now that this is not the case, and the
propagator for the fluctuation $\delta\gmn$ around flat space is not
affected by the higher derivative corrections to
Equation (\ref{bm-hdeom}). Namely, consider the solution for $\fmn$ in terms of
$\gmn$ given in Equation (\ref{pmf-sol}). Taking into account the fact that the
terms nonlinear in $P_{\mu\nu}(g)$ do not contribute to the linearized
equation because $P_{\mu\nu}(\eta)=0$, only two derivatives are acting
on $\delta\gmn$. It is then also clear that when this solution is
plugged into the $\fmn$ equation of motion, the resulting linearized
equation will contain at most two more derivatives acting on
$\delta\gmn$ coming from the Einstein tensor for $\fmn$. Hence, the
linearized equation for $\gmn$ around flat space is fourth order in
derivatives. \mbox{The fluctuation} equation obtained in this way is of
course the same as the linearized version of \mbox{Equation (\ref{bm-hdeom})}, in
which all four-derivative terms come from the Bach tensor. Hence, the
propagator around flat space is obtained from the Bach tensor alone
and is therefore identical to the propagator in conformal gravity. A
similar argument is valid for dS backgrounds, confirming the result
obtained in the bimetric formulation that a linear gauge symmetry is
present around these backgrounds.

Furthermore, in the small curvature limit, the
symmetry transformations of the bimetric fields $\gmn$ and $\fmn$
follow from the Weyl invariance of conformal gravity. Thus, to linear
order in curvatures and in the gauge parameter, and modulo coordinate
transformations, we get,
\be
\gmn'=\gmn +\phi\gmn\,,\qquad
\fmn'=\fmn -\frac{\phi}{\alpha^2}\gmn -\frac{1}{\beta_2 m^2}
\nabla^g_\mu\nabla^g_\nu \phi\,.
\label{transform}
\ee

Let us now consider the relation between dS backgrounds in
conformal gravity and in the PM bimetric theory. The CG equation
Equation \eqref{CGeom} admits dS solutions for which $R_{\mu\nu}(\bar g) =
\bar g_{\mu\nu}\Lambda$ (or $P_{\mu\nu}(\bar g)=\frac{1}{3}\bar
 g_{\mu\nu}\Lambda$). The
cosmological constant is not determined by the theory and can be
changed to any value by constant Weyl scalings of the metric, which
are symmetries of the theory.

This solution of CG is mapped to bimetric theory by
Equation \eqref{pmf-sol}. It turns out that on dS backgrounds, all higher
curvature corrections to Equation \eqref{pmf-sol} vanish, and the terms
explicitly written out give,
\be
\bar f_{\mu\nu}=\left(-\frac{1}{\alpha^2}+\frac{\Lambda}{3\beta_2 m^2}
 \right)\bar g_{\mu\nu}\,.
\ee

Writing this as $\bar f_{\mu\nu}=c^2\bar g_{\mu\nu}$, we obtain,
\be
\Lambda= m^2\left(\frac{1+c^2\alpha^2}{\alpha^2}\right) 3\beta_2\,,
\ee
which is indeed the expression for the cosmological constant in terms
of parameters of the PM bimetric theory. In this setup, $c$ is an
arbitrary gauge-dependent parameter. The flat space limit corresponds
to $c^2=-1/\alpha^2$. In the bimetric case, it is possible to express
the theory in terms of variables, such that all observables become
independent of $c$ \cite{Hassan:2012rq}.

Let us consider conformal gravity solutions away from dS spacetimes,
that is, when $R_{\mu\nu}(g)$ is not proportional to
$g_{\mu\nu}$. Such solutions, which are Bach flat, but not conformally
Einstein, are known to exist, see, e.g.,~\cite{Nurowski:2000cq, Liu:2013fna}. Then, from
Equation \eqref{pmf-sol}, it follows that in the bimetric setup, the field
$\fmn$ is not proportional to $\gmn$. In such cases, the bimetric
spectrum does not generically decompose into a massless and a massive
spin-2 field \cite{Hassan:2012wr}. Consequently, on the conformal
gravity side, too, the spectrum is not expected to follow such a
decomposition away from dS backgrounds.

\section{Discussions}

 \label{disc}
Our results are summarized in the Introduction section. Here, we would
like to first reemphasize \mbox{two issues}.
\begin{enumerate}
\renewcommand{\labelenumi}{(\theenumi)}
\item The analysis in this paper shows that the HR bimetric theory
 captures the essential features of higher derivative gravity action
 Equation \eqref{SHD2}, while at the same time avoiding the spin-2 ghost
 problem. The correspondence between the two theories found here is
 not a complete equivalence of equations of motion, but can still be
 used to generate higher derivative completions of the four-derivative
 gravity actions.
\item The equation of motion in the candidate PM bimetric theory at
 the four-derivative level was shown to coincide with the Bach
 equation of conformal gravity. While this result was motivated by
 the general correspondence between bimetric and HD gravity actions,
 it turns out to bypass the general correspondence and, in fact, was
 an equivalence at the level of equations of motion. As a result, it
 has genuine consequences for the bimetric PM proposal, as discussed
 in the paper.
\end{enumerate}

We also reiterate that the analysis of
\cite{Deser:2012qx,Deser:2013uy}, on acausal propagation and the absence
of a nonlinear PM symmetry in massive gravity, relies on the presence
of a non-dynamical metric and, as it stands, breaks down for the HR
bimetric theory. For this reason, we have emphasized the distinctions
between the HR bimetric models and the dRGT massive gravity models.

One may also consider alternative formulations and generalizations of
the PM bimetric model in terms of vielbeins. Introducing vielbeins
$e^a_{~\mu}$ and $\tilde e^a_{~\mu}$ for $\gmn$ and $\fmn$,
respectively, the square-root matrix can be expressed as
$(\sqrt{g^{-1}f})^\mu_{~\nu}=e^\mu_{~c} \tilde e^c_{~\nu}$, provided
the vielbeins satisfy the symmetrization condition, $\eta_{ac}\tilde
e^c_{~\rho}e^\rho_{~b}=e^{~\rho}_{a}\tilde e^{~c}_{\rho}\eta_{cb}$, or
in matrix form,
\be
\eta\tilde e e^{-1}=(\eta\tilde e e^{-1})^T \,.
\ee

In general, this condition can be implemented through a Lorentz
transformation, though there are caveats \cite{Deffayet:2012zc,
 Hassan:2014gta} to this. However, it has also been shown that the
bimetric theory expressed in terms of unconstrained vielbeins contains
the above symmetrization condition as part of its equations of \mbox{motion
\cite{Hinterbichler:2012cn}}, in the sense that this condition is
always a solution. This work also formulates more general
multivielbein theories that, with their equivalent metric formulations
\cite{Hassan:2012wt}, open the way for investigating PM theories with
more than one extra gauge invariance.

It would also be interesting to consider the conformal gravity
analogues of the PM bimetric theory in higher dimensions. It is known
that on adding Lanczos--Lovelock terms to the bimetric action, one can
obtain candidate PM bimetric theories in higher dimensions, in
particular for $d=6$ and $d=8$ \cite{Hassan:2012rq}. One then expects
that on eliminating $\fmn$, the corresponding equations of motion, to
lowest order in the curvature expansion, coincide with the equations
of the recently constructed conformal gravity theories in six
\cite{Bonora:1985cq, Metsaev:2010kp} and eight dimensions
\cite{Boulanger:2004zf}.

\vspace{6pt}

{Note}: Since this paper first appeared as a preprint, some
works have appeared claiming negative results for the possibility of a
nonlinear extension of the PM theory in various setups
\cite{Deser:2013gpa,Joung:2014aba, Garcia-Saenz:2014cwa}. However, these
results do not yet extend to the present bimetric setup. In
particular, \cite{Deser:2013gpa} claims that the relation between the
PM bimetric candidate and an extension of conformal gravity, as found
in this paper, is obtained only at the level of the action (as in
Section \ref{actionlevel}); hence, it cannot be treated as an
equivalence between the corresponding equations of motion. This claim
is astonishingly incorrect. \mbox{The authors} in \cite{Deser:2013gpa} have
chosen to ignore that this issue had already been explicitly raised
and addressed in the present paper, where, in Section \ref{eomlevel},
the relation is established at the level of bimetric equations of
motion. \mbox{This issue} was then re-emphasized when summarizing the results
in Section \ref{summary} and, again, when discussing them in Section
\ref{disc}.

On the other hand, \cite{Joung:2014aba} carries out an interesting
analysis starting with cubic interactions of a PM field, but as argued
in the Appendix of that paper, these cubic terms indeed vanish for the
PM bimetric candidate. Therefore, it is not obvious that, in spite of this,
the analysis of \cite{Joung:2014aba} would still apply to bimetric
theory. Arguments that rule out PM symmetry in bimetric theory, based
on results from massive gravity, have been dealt with in
\cite{Hassan:2014vja}. In a forthcoming work, we will show that the
transformation Equation (\ref{transform}) can be extended to at least six-derivative order
\cite{PMsym}.

\section*{Acknowledgements:} We would like to thank Shailesh Lal and
Bo Sundborg for discussions. 
\appendix

\section{Higher Derivative Treatment of Free Massive Spin-0 and Spin-2
 Fields}

To better understand the relation between bimetric theory and higher
derivative gravity and to resolve the ghost problem of the latter, we
first illustrate the procedure in a much simpler scalar field analogue
of the bimetric action. Then, we consider the linearized bimetric
theory to emphasize the parallel with the scalar field case.

\subsection{Higher Derivative Treatment of Massive Scalars}

Consider the theory of two scalar fields $\phi$ and $\psi$ with a
$\phi\psi$ mixing,
\be
S=\int d^dx\left[-\frac1{2}\p_\mu\phi\p^\mu\phi-\frac1{2}\p_\mu\psi\p^\mu
\psi-\frac{\mu^2}{2}(\phi+\psi)^2-V(\phi,\psi)
-\phi J_\phi-\psi J_\psi\right]\,.
\label{Ls}
\ee

It describes a pair of massless and massive scalars,
$\Phi_0=\phi-\psi$, $\Phi_m=\phi+\psi$, in a non-diagonal
basis. Conventionally, this theory is analyzed in terms of the mass
eigenstates. Here, we want to discuss two approaches to recast
Equation \eqref{Ls} as a higher derivative theory of a single field, with
emphasis on the ghost problem. However, for later reference, we first
review the usual discussion of ghosts in terms of the vacuum
persistence amplitude. For $V=0$, the field equations are,
\begin{align}
&\frac{\delta S}{\delta\phi}=0\quad \Rightarrow \quad
\square\phi-\mu^2(\phi+\psi)=J_\phi\,,\label{phi-eom}\\
&\frac{\delta S}{\delta\psi}=0\quad \Rightarrow \quad
\square\psi-\mu^2(\phi+\psi)=J_\psi\,,\label{psi-eom}
\end{align}

On adding and subtracting, they diagonalize to,
\be
\label{s-eigenstates}
\square\Phi_0 =J_0\,,\qquad
\left(\square-2\mu^2\right)\Phi_m=J_m\,,
\ee
with:
\be
\Phi_0=\phi-\psi\,,\qquad \Phi_m=\phi+\psi\,,\qquad
J_0=J_\phi-J_\psi\,,\qquad J_m=J_\phi+J_\psi\,.
\ee

Hence, Equation (\ref{Ls}) describes a massless scalar $\Phi_0$ and a massive
one $\Phi_m$ of mass $2\mu^2$. In particular, the solution for
$\phi=\frac{1}{2}(\Phi_0+\Phi_m)$ is,\footnote{We use the standard
 notation $(\square -m^2)^{-1}J(x)=\int d^dy\,G(x-y)J(y)$ with the
 Green's function $G(x-y)=(\square -m^2)^{-1}\delta(x-y)=-\int
 d^dk \frac{e^{ik(x-y)}}{k^2+m^2+i\epsilon}$.}
\be
\phi=\frac{1}{2}\Big[\square^{-1} J_0 +
(\square -2\mu^2)^{-1}J_m + \Phi_0^{\hom} + \Phi_m^{\hom}\Big]\,,
\label{phi-sol}
\ee
where $\Phi_{0,m}^{\hom}$ are the homogeneous solutions of
Equation (\ref{s-eigenstates}). From Equation (\ref{Ls}), on completing squares, one
reads off the vacuum persistence amplitude,
\be
A=\int{\cal D}\phi\, e^{iS}=
\exp\left(-\frac{i}{4}\int d^d x\left[J_0(x)\square^{-1} J_0(x)
+ J_m(x)(\square -2\mu^2)^{-1} J_m(x)\right]\right) < 1\,.
\label{vpa}
\ee

The propagators $i\square^{-1}$ and $i(\square -2\mu^2)^{-1}$ have
positive residues at the corresponding mass poles, which results in
$A<1$, meaning that the theory is ghost free. In other words, the
vacuum has decayed only into positive norm states due to interaction
with the sources. In a theory that contains ghosts, it becomes possible
to get $A>1$, implying a breakdown of unitarity and signalling vacuum
decay into negative norm states. In the following, we will use this
criterion to check for ghosts in free higher derivative theories where
$A$ can be computed.

The potential $V(\phi,\psi)$ provides the interactions between the two
mass eigenstates. At low energies, the massive mode $\Phi_m$ can be
integrated out. In particular, for $V=0$, this gives a Lagrangian for
the free massless field $\Phi_0$ with no knowledge of the massive
mode. The analysis below does not involve integrating out fields in
this way. Let us now consider the two approaches to obtaining higher
derivative equations from Equation \eqref{Ls}.

\subsection{First Approach: The Equivalent Higher Derivative
 Equations}

Instead of diagonalizing the equations of motion Equations (\ref{phi-eom}) and
(\ref{psi-eom}), it is instructive to rewrite them as an equivalent
higher derivative equation. The purpose is to illustrate that, given a
higher derivative equation, the appearance of ghosts is not a forgone
conclusion, but depends on the structure of the source terms and on the
choice of action.

Let us eliminate $\psi$ between the equations of motion. The
$\phi$-equation, Equation (\ref{phi-eom}), is algebraic in $\psi$ and solves to,
\be
\label{psisol}
\psi = -\phi+\frac1{\mu^2}\left(\square\phi-J_\phi\right)\,.
\ee

Such algebraic solutions can be obtained even for a non-zero
$V(\phi,\psi)$, at least perturbatively in $1/\mu^2$.
Inserting this in Equation \eqref{psi-eom} gives a higher derivative
equation,\footnote{The same result follows on first solving
 Equation \eqref{psi-eom} for $\psi$ and substituting the outcome in
 Equation \eqref{phi-eom}.}
\be
\label{phieq1}
\square\left(\square-2\mu^2\right)\phi= \frac{1}{2}\left[(\square
-2\mu^2) J_0+\square J_m \right]\,.
\ee

The specific source structure in Equation (\ref{phieq1}) and the choice of
action leading to it are crucial for the absence of a ghost. Had we
ignored the sources in the above derivation and instead introduced a
generic source coupling at the end, we would obtain
$\square\left(\square-2\mu^2\right)\phi=j$, corresponding to an action
$\int d^dx[\tfrac{1}{2}\phi\square\left(\square-2\mu^2\right)\phi
 -j\phi]$. This gives the correct mass poles, but implies that the
massive mode is a ghost.\footnote{In the naive propagator for this
 theory, $\frac{1}{k^2(k^2+m^2)}=\frac{1}{m^2} (\frac{1}{k^2}-
 \frac{1}{k^2+m^2})$, the residues of the massless and massive poles
 have different signs, so in the analogue of Equation \eqref{vpa}, $A$ is not
 always bounded by unity.} Of course, it is not obvious why two
propagating modes of different masses should couple to the same source
with the same strength.

However, with the source structure of Equation (\ref{phieq1}), one gets the
correct solution Equation (\ref{phi-sol}). \mbox{To check} for the presence of ghosts,
one needs to specify an action. The obvious procedure is to integrate
out $\psi$ in Equation \eqref{Ls} (by completing the squares for $\psi$ or,
equivalently, eliminating $\psi$ through the solution of {its own}
equation of motion, and not Equation (\ref{psisol})). Then, for $V=0$, one
obtains the bi-local higher derivative action for $\phi$,

\begin{align}
S_{bl}=\frac{1}{2}\int d^dx d^dy\bigg[&\phi(x) G(x-y)\Big[
\square_y(\square_y -2\mu^2)\phi(y)- (\square_y-2\mu^2)J_o(y)
+\square_y J_m(y)\Big]\nn\\
&\qquad -\frac{1}{4}[J_m(x)-J_0(x)] G(x-y)[J_m(y)
-J_0(y)]\bigg]\,.
\end{align}

By construction, this has Equation \eqref{phieq1} as the equation of motion
and gives the vacuum persistence amplitude Equation \eqref{vpa}, consistent
with the absence of ghosts. Although the $J-J$ term in the second
line does not contribute to the equation of motion, it is needed
to get the correct ghost-free amplitude. Its pretense simply indicates
that the interactions of the $\psi$ field cannot all be encoded in the
structure of the $\phi$ action alone.
To summarize, the above illustrates that a higher-derivative equation,
like \mbox{$\square(\square-2\mu^2)\phi=0$}, which normally implies a ghost,
could actually be ghost free with the appropriate action and \mbox{source
couplings}.

\subsection{Second Approach: A More General Higher Derivative Action}

Now, instead of using the algebraic solution $\psi(\phi,J_\phi)$ to
eliminate $\psi$ from the equations of motion (first approach), we use
it in the action to get $S[\phi,\psi(\phi,J_\phi)]$. This gives a
six-derivative equation of motion for $\phi$, which is more general and
contains the solutions of Equation \eqref{phieq1} as a subsector. \mbox{The analogue}
of this treatment, when applied to bimetric theory, reproduces the
higher derivative gravity theories. \mbox{The discussion} here is intended to
shed light on the emergence of the spin-2 ghost in higher \mbox{derivative
gravity}.

Using the solution Equation (\ref{psisol}) to eliminate $\psi$ in the action
Equation (\ref{Ls}) gives a six-derivative action,
\begin{align}
S[\phi,\psi(\phi,J_\phi)]=\frac1{2\mu^4}\int\td^dx\biggl[&\phi\,\square
\left(\square-\mu^2\right)\left(\square-2\mu^2\right)\phi
-2\,\phi\left(\square-\mu^2\right)\left(\left(\square-
\mu^2\right)J_\phi+\mu^2 J_\psi\right)\nn\\
&\qquad\qquad+J_\phi(\square-\mu^2)J_\phi
+2\mu^2 J_\phi J_\psi\biggr]\,.
\label{hd2}
\end{align}

The corresponding equation of motion is,
\be
\frac{\delta S}{\delta\phi(x)}\bigg\vert_{\psi} +
\int d^dy \frac{\delta\psi(y)}{\delta\phi(x)} \,
\frac{\delta S}{\delta\psi(y)}\bigg\vert_{\phi}=0\,.
\ee

The first term vanishes by virtue of Equation (\ref{psisol}).
Hence, one gets, $\int d^dy\frac{\delta\psi(y)}{\delta\phi(x)}
\frac{\delta S}{\delta\psi(y)}\big\vert_{\phi}=0$, which differs from
$\frac{\delta S}{\delta\psi}=0$ by the extra operator,
$\frac{\delta\psi(y)}{\delta\phi(x)}=\frac{1}{\mu^2}(\square_y-\mu^2)
\delta(x-y)$, as follows form Equation \eqref{psisol}. This gives the higher
derivative equation,
\be
\label{phieq2}
\left(\square-\mu^2\right)\left[\square\left(\square-2\mu^2\right)\phi
-\frac{1}{2}\left((\square -2\mu^2) J_0+\square J_m \right)\right]
=0\,.
\ee

The terms within square brackets are the same as in Equation \eqref{phieq1}.
Had we ignored the sources from the beginning and then inserted
a generic source coupling at the end, we would obtain
\mbox{$\square(\square-\mu^2)(\square-2\mu^2)\phi=j$}, corresponding to an
action $\int d^4x \frac{1}{2}\phi\square(\square-\mu^2)(\square-
2\mu^2)\phi-j\phi$. \mbox{This would} give a theory with three mass poles and
a ghost mode. However, for the correct source structure,
\mbox{Equation \eqref{phieq2}} is equivalent to,
\be
\square\left(\square-2\mu^2\right)\phi
-\frac{1}{2}\left((\square -2\mu^2) J_0+\square J_m \right)
=\chi^{\hom}\,,
\ee
where $\chi^{\hom}$ solves $(\square-\mu^2)\chi^{\hom}=0$ and needs to
be specified through two boundary conditions. \mbox{The boundary} conditions
that give $\chi^{\hom}=0$ specify the subsector of Equation \eqref{hd2} that is
equivalent to Equation \eqref{Ls}.

In general, the complete solution becomes,
\be
\phi=\frac{1}{2}\Big[\square^{-1} J_0 +
(\square -2\mu^2)^{-1}J_m\Big]\,
+\square^{-1}\left(\square-2\mu^2\right)^{-1}\chi^{\hom}
+\frac{1}{2}\Big[\Phi_0^{\hom} + \Phi_m^{\hom}\Big]\,.
\label{phi-sol-chi}
\ee

A non-vanishing $\chi^{\hom}$ appears as an arbitrariness in the sources
$J_0$ and $J_m$, but even then, only the two original mass poles of
Equation \eqref{Ls} contribute to the solution.

To address the ghost issue, one can again compute the vacuum
persistence amplitude $\int{\cal D}\phi e^{iS}$ by completing the
squares for $\phi$ in the action Equation \eqref{hd2}. The answer turns out to
be exactly the same as Equation \eqref{vpa}. In particular, there is no
contribution from $\chi^{\hom}$. Hence, the complete untruncated higher
derivative theory Equation \eqref{hd2} does not propagate ghosts.

\subsection{Truncation to a Four Derivative Theory}

In the action Equation \eqref{hd2}, the highest derivative terms come with
$1/\mu^4$. Dropping these results in a four-derivative action for $\phi$
with mass poles at $m^2=0$ and $m^2=2\mu^2/3$. On completing
the squares, one can see that the massless mode is healthy, while the
massive mode is now a ghost. Thus, the truncation has shifted the
massive pole away from $2\mu^2$ and turned it into a ghost. Of course,
in this example, we know that the theory cannot be trusted near the
massive pole, which is close to the truncation scale. Near the massive
pole, higher derivative corrections become important, and including
these both shifts the mass pole to its correct value, as well as
eliminates the ghost.

The analogue of Equation \eqref{hd2} in the bimetric setup is given as an
expansion in powers of derivatives, suppressed by the mass
parameter. When truncated to four-derivative terms, the massive spin-2
field becomes a ghost with a shifted mass, as an artifact of the
truncation. The lesson is that the spin-2 ghost in HD gravity must be
understood in a similar way, although in the presence of interactions,
the complete HD theory will not have a finite number of terms.

\subsection{Higher Derivative Treatment of Linearized Bimetric Theory}

Let us now repeat this procedure for spin-2 fluctuations on a curved
background $\bgmn$. The notation here simplifies and streamlines the
manipulations. Indices are raised using $\bar g$. Our starting point
is the quadratic Fierz--Pauli action with linear interaction given by,
\be
S=\int\td^dx\sqrt{\bar g}\biggr[
\frac1{2}h_{\mu\nu}D^{\mu\nu\rho\sigma}(\mu^2,\Lambda)h_{\rho\sigma}
+\frac1{2}{\tilde h}_{\mu\nu}D^{\mu\nu\rho\sigma}(\tilde\mu^2,\tilde\Lambda){\tilde h}_{\rho\sigma}
+\lambda\, h_{\mu\nu}G^{\mu\nu\rho\sigma}_{(1)}{\tilde h}_{\rho\sigma}
-h_{\mu\nu}T^{\mu\nu}-{\tilde h}_{\mu\nu}\tilde T^{\mu\nu}
\biggl]\,,
\ee
where we have defined,
\be
D^{\mu\nu\rho\sigma}(\mu^2,\Lambda)=\mathcal{E}^{\mu\nu\rho\sigma}
+\frac{\mu^2}{2}G^{\mu\nu\rho\sigma}_{(1)}
-\frac{\Lambda}{d-2}G^{\mu\nu\rho\sigma}_{(1/2)}\,,
\ee
in terms of the linear Einstein operator
$\mathcal{E}^{\mu\nu\rho\sigma}$ and the generalized DeWitt metrics,
which we define~through,
\be
G^{\mu\nu\rho\sigma}_{(\xi)}=\bar g^{\mu\nu}\bar g^{\rho\sigma}-\xi\,\bar g^{\mu(\rho}\bar g^{\sigma)\nu}\,.
\ee

The field equations are given by,
\be\label{hEq}
D^{\mu\nu\rho\sigma}(\mu^2,\Lambda)h_{\rho\sigma}
+\lambda\, G^{\mu\nu\rho\sigma}_{(1)}{\tilde h}_{\rho\sigma}-T^{\mu\nu}=0\,,
\ee
\be\label{fEq}
D^{\mu\nu\rho\sigma}(\tilde\mu^2,\tilde\Lambda){\tilde h}_{\rho\sigma}
+\lambda\, G^{\mu\nu\rho\sigma}_{(1)}h_{\rho\sigma}-\tilde T^{\mu\nu}=0\,,
\ee

Restricting now to $\tilde\Lambda=\Lambda$ and
$\tilde\mu^2=2\lambda=\mu^2$, we can again simply add and subtract
these equations to get,
\be
D^{\mu\nu\rho\sigma}(2\mu^2,\Lambda)\left(h_{\rho\sigma}+{\tilde h}_{\rho\sigma}\right)
-\left(T^{\mu\nu}+\tilde T^{\mu\nu}\right)=0\,,
\ee
\be
D^{\mu\nu\rho\sigma}(0,\Lambda)\left(h_{\rho\sigma}-{\tilde h}_{\rho\sigma}\right)
-\left(T^{\mu\nu}-\tilde T^{\mu\nu}\right)=0\,.
\ee

These are the equations of a massive spin-2 field
$h^+_{\mu\nu}=h_{\mu\nu}+{\tilde h}_{\mu\nu}$ with a mass $2\mu^2$ and
a massless spin-2 field $h^-_{\mu\nu}=h_{\mu\nu}-{\tilde h}_{\mu\nu}$
on a curved background. We can solve for ${\tilde h}_{\mu\nu}$ in
Equation \eqref{hEq}, to get,
\be\label{fsol1}
G^{\mu\nu\rho\sigma}_{(1)}{\tilde h}_{\rho\sigma}=-\frac{2}{\mu^2}\left[
D^{\mu\nu\rho\sigma}(\mu^2,\Lambda)h_{\rho\sigma}-T^{\mu\nu}\right]\,,
\ee
and subsequently,
\be\label{fsol2}
{\tilde h}_{\mu\nu}=\frac{2}{\mu^2}\left[D_{\mu\nu}^{\ph{\mu\nu}\rho\sigma}(\mu^2,\Lambda)h_{\rho\sigma}-T_{\mu\nu}
-\frac{1}{d-1}\bgmn\left(D^{\alpha\ph\alpha\rho\sigma}_{\ph\alpha\alpha}(\mu^2,
\Lambda)h_{\rho\sigma}-T\right)\right]\,.
\ee

Inserting this solution in Equation \eqref{fEq} and using some basic
manipulations (see the end of this section) of the DeWitt metrics, one
obtains the higher derivative equation,
\be\label{dheq}
D^{\mu\nu\alpha\beta}(0,\Lambda)G^{(d-1)}_{\alpha\beta\rho\sigma}D^{\rho\sigma\lambda\kappa}(2\mu^2,\Lambda)
h_{\lambda\kappa}-\left(D^{\mu\nu\alpha\beta}(\mu^2,\Lambda)G^{(d-1)}_{\alpha\beta\rho\sigma}T^{\rho\sigma}
-\frac{d-1}{2}\mu^2\,\tilde T^{\mu\nu}\right)=0\,.
\ee

Again, the structure of the higher derivative operator hitting
$h_{\lambda\kappa}$ is the composite of a zero mass operator and a
massive operator for a mass $2\mu^2$. The contraction of these
operators is of course a bit more complex now, involving a DeWitt
metric. In fact, this is the inverse of $G_{(1)}$, in the sense that,
\be
G^{\mu\nu\rho\sigma}_{(1)}G^{(d-1)}_{\alpha\beta\rho\sigma}=\frac{d-1}{2}\left(\delta^\mu_\alpha\delta^\nu_\beta
+\delta^\mu_\beta\delta^\nu_\alpha\right)\,.
\ee

In the source contribution of Equation \eqref{dheq}, one of the sources is
contracted with an operator $\sim\delta {\tilde h}^{\mu\nu}/\delta
h_{\rho\sigma}$. All of this is in complete analogy to the scalar
field example and generalizes that discussion in a very
straightforward way, the only complication being the metric of
contraction. \mbox{The insertion} of the solution Equations \eqref{fsol1} and
 \eqref{fsol2} into the action also proceeds in an analogue fashion and
again results in an operator $D(\mu^2,\Lambda)$ contracting
Equation \eqref{dheq} with the DeWitt metric $G^{(d-1)}$. \mbox{This makes} it obvious
that the implications of absence of a ghost for the untruncated theory
go through just as in the scalar field example, and we do not provide
the details of this here.

In order to check the details at the level of the action, it is useful
to observe that,
\be
G^{\mu\nu\alpha\beta}_{(1)}G_{(1)\alpha\beta}^{\ph{(1)\alpha\beta}\rho\sigma}
+G^{\mu\nu\rho\sigma}_{(1)}-\frac1{d-1}G^{\mu\nu\alpha}_{(1)\ph\mu\alpha}
G^{\beta\ph{(1)}\rho\sigma}_{(1)\beta}=0\,,
\ee
and,
\be
G^{~\alpha\ph\alpha\rho\sigma}_{(1)\alpha}=G^{\rho\sigma\alpha}_{(1)~\alpha}=(d-1)\,\bar g^{\rho\sigma}\,,
\ee
together with (for any $D$),
\be
D^{\alpha\beta\rho\sigma}G^{\mu\nu}_{(1)\alpha\beta}
=D^{\alpha\ph\alpha\rho\sigma}_{\ph\alpha\alpha}\bar g^{\mu\nu}
-D^{\mu\nu\rho\sigma}\,,\qquad
D^{\mu\nu}_{\ph{\mu\nu}\alpha\beta}G_{(1)}^{\alpha\beta\rho\sigma}
=D^{\mu\nu\alpha}_{\ph{\mu\nu\alpha}\alpha}\bar g^{\rho\sigma}
-D^{\mu\nu\rho\sigma}\,.
\ee

These results can be used to show that,
\begin{align}
&D^{\mu\nu\alpha\beta}(\mu^2,\Lambda)D_{\alpha\beta}^{\ph{\alpha\beta}\rho\sigma}(\mu^2,\Lambda)
-\frac{1}{d-1}D^{\mu\nu\alpha}_{\ph{\mu\nu\alpha}\alpha}(\mu^2,\Lambda)
D^{\beta\ph\beta\rho\sigma}_{\ph\beta\beta}(\mu^2,\Lambda)+\frac{\mu^4}{4}G^{\mu\nu\rho\sigma}_{(1)}\nn\\
&=D^{\mu\nu\alpha\beta}(0,\Lambda)D_{\alpha\beta}^{\ph{\alpha\beta}\rho\sigma}(2\mu^2,\Lambda)
-\frac{1}{d-1}D^{\mu\nu\alpha}_{\ph{\mu\nu\alpha}\alpha}(0,\Lambda)
D^{\beta\ph\beta\rho\sigma}_{\ph\beta\beta}(2\mu^2,\Lambda)\,.
\end{align}

{Together with the solutions Equations \eqref{fsol1} and \eqref{fsol2}, these
relations make the {manipulations straightforward}.}

\section{The General Perturbative Solution of the $\gmn$ Equation for
$\fmn$}

For general $\beta_n$, finding an exact solution for $S=\sqrt{g^{-1}f}$
is not feasible.\footnote{In principle, exact solutions can be
 obtained implicitly by considering higher powers of the
 $\gmn$-equation when written in the form $R^\mu_{~\nu}
 =\sum_r q_r (S^r)^\mu_{~\nu}$, where $q_r$ are functions of
 $\beta_n$ and $e_n(S)$. On tracing and expressing $\Tr(S^r)$ for
 $r>d$ in terms $\Tr(S^n)$ with $n\leq d$, one can obtain enough
 equations to determine all $e_n(S)$. \mbox{Then, the} $g$-equation can be
 used to determine $S$ in terms of the curvatures.} However, one can find
a perturbative solution in powers of $\frac{1}{m^2}$. The equation of
motion for $\gmn$ in terms of its Schouten tensor $\pmn$ becomes,
\be\label{Aeomggen}
\Tr P\,\delta^\mu_\nu-\Pmn=m^2\sum_{n=0}^{d-1}(-1)^n\beta_n
\mathbb{Y}_{(n)\nu}^{\mu}(S)\,,
\ee
where the matrices $\mathbb{Y}_{(n)}(S)$ are defined in
Equation (\ref{ydef}), and here and in the following, we raise and lower indices
with $\gmn$. Our goal is to solve the Equation (\ref{Aeomggen}) for
$\Smn$ perturbatively in $\Pmn/m^2$. \mbox{Since the} solution for $\Smn$ is
a polynomial in~$\Pmn$, the resulting action for $\gmn$ will
be a higher curvature theory with coefficients given in terms of the
parameters of the bimetric model. In order to find the perturbative
solution for $\Smn$, consider the general ansatz,
\begin{align}
\label{Aansatzs}
\Smn=a\delta^\mu_\nu &+\frac{1}{m^2}\left(b_1\Pmn+b_2\Tr
P\,\delta^\mu_\nu \right)\nn\\
&+\frac{1}{m^4}\left(c_1{\Pmn}^2+c_2\Pmn\Tr P+c_3\Tr (P^2)\delta^\mu_\nu
+c_4(\Tr P)^2\delta^\mu_\nu\right)~+~\mathcal{O}(m^{-6})\,.
\end{align}

When we plug this into Equation (\ref{Aeomggen}), the coefficients of different
terms $\delta^\mu_\nu$, $\Pmn$, $\delta^\mu_\nu\Tr P$, ${\Pmn}^2$,
$\hdots$ have to vanish
separately. This will determine all of the coefficients $a, b_i, c_i$ in
the ansatz for $\Smn$.

To simplify the calculation, define $M^\mu_{~\nu}\equiv
\Smn/a-\delta^\mu_\nu$ or, $\Smn=a(\delta^\mu_\nu+M^\mu_{~\nu})$.
Then, using,
\beqn
e_k(S)=a^k\sum_{m=0}^k{d-m \choose k-m}e_m(M)\,,\qquad
(S^{n-k})^\mu_{~\nu}=a^{n-k}\sum_{r=0}^{n-k}{n-k\choose r}(M^r)^\mu_{~\nu}\,,
\eeqn
the equations of motion can be written as,
\beqn\label{Aeombin}
\Tr P\,\delta^\mu_\nu-\Pmn=m^2\sum_{n=0}^{d-1}\beta_na^{n}\sum_{k=0}^n
\sum_{m=0}^k\sum_{r=0}^{n-k}(-1)^{n+k}{d-m \choose k-m}{n-k\choose r}
e_m(M)\,(M^r)^\mu_{~\nu}\,.
\eeqn

Since $M^\mu_{~\nu}$ starts with $m^{-2}$, a power of $M^l$ will
contribute to orders $m^{-2l}$ and higher. In the following, we work out
the first three orders explicitly. To simplify the notation,
we use the sums $s_k$ defined in Equation \eqref{sums} that will be frequently
encountered below.

\vspace{6pt}
\noindent{\it Zeroth order:} The equation of motion at
$\mathcal{O}(m^0)$ is obtained from terms with the last two sums in
Equation~(\ref{Aeombin}) restricted to $(r,m)=(0,0)$. This gives
$s_0\equiv\sum_{n=0}^{d-1}{d-1 \choose n}\beta_n a^n=0$,
Equation \eqref{eq1or}.
This is a polynomial equation in $a$ that determines it in terms of
the $\beta_n$.

\noindent{\it First order:} At $\mathcal{O}(m^{-2})$, terms with
$(r,m)=(1,0)$ and $(r,m)=(0,1)$ contribute in Equation (\ref{Aeombin}). Using
simple identities for the binomial coefficients, we obtain for this
order,
\beqn\label{app: Pfirst}
\Tr P\,\delta^\mu_\nu-\Pmn=\frac{1}{a}\Big[\big((d-1)b_2+b_1\big)\Tr
 P\,\delta^\mu_\nu-b_1\Pmn\Big]s_1\,,
\eeqn
where $s_1$ is defined in Equation (\ref{sums}). Comparing the coefficients of
$\Tr P\,\delta^\mu_\nu$ and $\Pmn$ separately, we find the following
equation that determines $b_1$ and $b_2$ in terms of the $\beta_n$.
\be\label{Aeq2or}
b_2=0\,,\qquad b_1=\frac{a}{s_1}\,.
\ee

\noindent{\it Second order:} At order $\mathcal{O}(m^{-4})$, we can
have contributions from the following combinations of $(r,m)$,
\be
(1,0)\,,\quad (2,0)\,,\quad (1,1)\,,\quad (0,1)\,,\quad (0,2)\,.
\ee

Demanding the coefficients of the different terms at this order to
vanish separately then gives the following system of equations for the
coefficients $c_i$ in the ansatz Equation (\ref{Aansatzs}),
\beqn\label{Asys}
s_1c_1-\frac{b_1^2s_2}{a}&=&0\,,\qquad
s_1c_2+\frac{b_1^2s_2}{a}~=~0\,,\nn\\
s_1(c_1+dc_3)-s_1c_3-\frac{b_1^2s_2}{2a}&=&0\,,\qquad
s_1(c_2+dc_4)-s_1c_4+\frac{b_1^2s_2}{2a}~=~0\,.
\eeqn

The solution is easily found to be,
\be
c_1=-c_2=\frac{b_1^2s_2}{as_1}=\frac{as_2}{s^3_1}\,,\qquad c_3=-c_4=
\frac{c_2}{2(d-1)}=-\frac{as_2}{2(d-1)s^3_1} \,,
\ee
where we have also used the solution for $b_1$ given in
Equation (\ref{eq2or}). Putting everything together, the solution for $\Smn$ to
$\mathcal{O}(m^{-4})$ becomes Equation \eqref{ssol}. The procedure can
straightforwardly be continued to compute the solution for $\Smn$ to
any order in $m^{-2}$.



\begin{thebibliography}{999}
\bibitem{Hassan:2011zd}
 Hassan, S.F.; Rosen, R.A.
 Bimetric Gravity from Ghost-free Massive Gravity.
 \emph{J. High Energy Phys.} {\bf 2012}, \emph{2012}, 126. 

\bibitem{Hassan:2011ea}
 Hassan, S.F.; Rosen, R.A.
 Confirmation of the Secondary Constraint and Absence of Ghost in
 Massive Gravity and Bimetric Gravity.
 \emph{J. High Energy Phys.} {\bf 2012}, \emph{2012}, 123.
\bibitem{Hassan:2012gz}
 Hassan, S.F.; Schmidt-May, A.; von Strauss, M.
 On Partially Massless Bimetric Gravity.
 \emph{\mbox{Phys. Lett. B}} \textbf{2012}, \emph{726}, 834--838. 
\bibitem{Hassan:2012rq}
 Hassan, S.F.; Schmidt-May, A.; von Strauss, M.
 Bimetric Theory and Partial Masslessness with Lanczos-Lovelock Terms in Arbitrary Dimensions. \emph{Class. Quant. Grav.} \textbf{2012}, \emph{30}, 184010

\bibitem{Stelle:1976gc}
 Stelle, K.S.
 Renormalization of Higher Derivative Quantum Gravity.
 \emph{Phys. Rev. D} {\bf 1977}, \emph{16}, 953--969.
\bibitem{Stelle:1977ry}
 Stelle, K.S.
 Classical Gravity with Higher Derivatives.
 \emph{Gen. Rel. Grav.} {\bf 1978}, \emph{9}, 353--371.
\bibitem{Bergshoeff:2009hq}
 Bergshoeff, E.A.; Hohm, O.; Townsend, P.K.
 Massive Gravity in Three Dimensions.
 \emph{\mbox{Phys. Rev. Lett.}} {\bf 2009}, \emph{102}, 201301.
\bibitem{Ohta:2011rv}
 Ohta, N.
 A Complete Classification of Higher Derivative Gravity in 3D and Criticality in 4D.
 \emph{\mbox{Class. Quant. Grav.}} {\bf 2012}, \emph{29}, 015002.

\bibitem{Kleinschmidt:2012rs}
 Kleinschmidt, A.; Nutma, T.; Virmani, A.
 On unitary subsectors of polycritical gravities.
 \emph{\mbox{Gen. Rel. Grav.}} {\bf 2013}, \emph{45}, 727--749.

 \bibitem{Eliezer:1989cr}
 Eliezer, D.A.; Woodard, R.P.
 The Problem of Nonlocality in String Theory.
 \emph{Nucl.\ Phys.\ B} {\bf 1989}, \emph{325}, 389--469.
 \bibitem{Simon:1990ic}
 Simon, J.Z.
 Higher Derivative Lagrangians, Nonlocality, Problems And Solutions.
 \emph{Phys.\ Rev.\ D} {\bf 1990}, \emph{41}, 3720--3733.
 \bibitem{Biswas:2005qr}
 Biswas, T.; Mazumdar, A.; Siegel, W.
 Bouncing universes in string-inspired gravity.
 \emph{J. Cosmol. Astropart. Phys.} {\bf 2006}, doi:10.1088/1475-7516/2006/03/009.

\bibitem{Biswas:2011ar}
 Biswas, T.; Gerwick, E.; Koivisto, T.; Mazumdar, A.
 Towards singularity and ghost free theories of gravity.
 \emph{Phys. Rev. Lett.} {\bf 2012}, \emph{108}, 031101.

\bibitem{Biswas:2012bp}
 Biswas, T.; Koshelev, A.S.; Mazumdar, A.; Vernov, S.Y.
 Stable bounce and inflation in non-local higher derivative cosmology.
 \emph{J. Cosmol. Astropart. Phys.} {\bf 2012}, doi:10.1088/1475-7516/\linebreak 2012/08/024.

\bibitem{Nojiri:2012zu}
 Nojiri, S.I.; Odintsov, S.D.
 Ghost-free $F(R)$ bigravity and accelerating cosmology.
 \emph{Phys.\ Lett.\ B} {\bf 2012},  doi:10.1016/j.physletb.2012.08.049.

\bibitem{bach}
 Bach, R.
Zur Weylschen Relativit\"atstheorie und der Weylschen Erweiterung des Kr\"ummungsbegriffs.
\emph{Math. Zeitschr.} {\bf 1921}, \emph{9}, 110--135. (In German)
\bibitem{Kaku:1977pa}
 Kaku, M.; Townsend, P.K.; van Nieuwenhuizen, P.
 Gauge Theory of the Conformal and Superconformal Group.
 \emph{Phys. Lett. B} {\bf 1977}, \emph{69}, 304--308.
\bibitem{Fradkin:1981iu}
 Fradkin, E.S.; Tseytlin, A.A.
 Renormalizable asymptotically free quantum theory of gravity.
 \emph{\mbox{Nucl. Phys. B}} {\bf 1982}, \emph{201}, 469--491.
\bibitem{Lee:1982cp}
 Lee, S.C.; van Nieuwenhuizen, P.
 Counting of States In Higher Derivative Field Theories.
 \emph{\mbox{Phys. Rev. D}} {\bf 1982}, \emph{26}, 934--937.
\bibitem{Riegert:1984hf}
 Riegert, R.J.
 The Particle Content Of Linearized Conformal Gravity.
 \emph{Phys.\ Lett.\ A} {\bf 1984}, \emph{105}, 110--112.
\bibitem{Maldacena:2011mk}
 Maldacena, J.
 Einstein Gravity from Conformal Gravity. \textbf{2011},
 arXiv:1105.5632 [hep-th].
\bibitem{Lu:2011ks}
 Lu, H.; Pang, Y.; Pope, C.N.
 Conformal Gravity and Extensions of Critical Gravity.
 \emph{Phys. Rev. D} {\bf 2011}, \emph{84}, 064001.

\bibitem{Lu:2013hx}
 Lu, H.; Pang, Y.; Pope, C.N.
 Black Holes in Six-dimensional Conformal Gravity. 	\emph{Phys. Rev. D} \textbf{2013}, \emph{87}, 104013.


\bibitem{Metsaev:2007rw}
 Metsaev, R.R.
 Ordinary-derivative formulation of conformal totally symmetric arbitrary spin bosonic fields.
 \emph{J. High Energy Phys.} {\bf 2012}, \emph{2012}, 062.
\bibitem{Mannheim:2011ds}
 Mannheim, P.D.
 Making the Case for Conformal Gravity.
 \emph{Found. Phys.} {\bf 2012}, \emph{42}, 388--420.

\bibitem{Schmidt:2006jt}
 Schmidt, H.J.
 Fourth order gravity: Equations, history, and applications to cosmology.
 \emph{Int. J. Geom. Meth. Mod. Phys.} {\bf 2007}, \emph{4}, 209--248.

\bibitem{Alexandrov:2012yv}
 Alexandrov, S.; Krasnov, K.; Speziale, S.
Chiral description of ghost-free massive gravity. \emph{J. High Energy Phys.} \textbf{2013}, \emph{2013}, 068.

\bibitem{Soloviev:2013mia}
 Soloviev, V.O.; Tchichikina, M.V.
 Bigravity in Kuchar's Hamiltonian formalism: The special case. \emph{Phys. Rev. D} \textbf{2013}, \emph{88}, 084026.


\bibitem{Hassan:2012wr}
 Hassan, S.F.; Schmidt-May, A.; von Strauss, M.
 On Consistent Theories of Massive Spin-2 Fields Coupled to Gravity. \emph{J. High Energy Phys.} \textbf{2013}, \emph{2013}, 86.
\bibitem{Volkov:2011an}
 Volkov, M.S.
 Cosmological solutions with massive gravitons in the bigravity theory.
 \emph{J. High \mbox{Energy Phys.}} {\bf 2012}, \emph{2012}, 035.

\bibitem{vonStrauss:2011mq}
 Von Strauss, M.; Schmidt-May, A.; Enander, J.; Mortsell, E.; Hassan, S.F.
 Cosmological Solutions in Bimetric Gravity and their Observational Tests.
 \emph{J. Cosmol. Astropart. Phys.} {\bf 2012},  doi:10.1088/1475-7516/2012/03/042.

\bibitem{Comelli:2011zm}
 Comelli, D.; Crisostomi, M.; Nesti, F.; Pilo, L.
 FRW Cosmology in ghost-free Massive Gravity.
 \emph{\mbox{J. High Energy Phys.}} {\bf 2012}, doi:10.1007/JHEP03(2012)067.

\bibitem{Berg:2012kn}
 Berg, M.; Buchberger, I.; Enander, J.; Mortsell, E.; Sjors, S.
 Growth Histories in Bimetric Massive Gravity. \emph{J. Cosmol. Astropart. Phys.} \textbf{2012},  doi:10.1088/1475-7516/2012/12/021.

\bibitem{Park:2012cq}
 Park, M.; Sorbo, L.
 Vacua and instantons of ghost-free massive gravity. \emph{Phys. Rev. D} \textbf{2013}, \emph{87},~024041.

 \bibitem{Sakakihara:2012iq}
 Sakakihara, Y.; Soda, J.; Takahashi, T.
 On Cosmic No-hair in Bimetric Gravity and the Higuchi Bound. \emph{Prog. Theor. Exp. Phys.} \textbf{2013}, \emph{2013}, 033E02.

\bibitem{Akrami:2012vf}
 Akrami, Y.; Koivisto, T.S.; Sandstad, M.
 Accelerated expansion from ghost-free bigravity: a statistical analysis with improved generality.
 \emph{J. High Energy Phys.} {\bf 2013}, \emph{2013}, 99.

\bibitem{Capozziello:2012re}
 Capozziello, S.; Martin-Moruno, P.
 Bounces, turnarounds and singularities in bimetric gravity.
 \emph{Phys. Lett. B} {\bf 2013}, \emph{719}, 14--17.

\bibitem{Mohseni:2012ug}
 Mohseni, M.
 Gravitational Waves in Ghost Free Bimetric Gravity.
 \emph{J. Cosmol. Astropart. Phys.} {\bf 2012},  doi:10.1088/1475-7516/2012/11/023.

\bibitem{Baccetti:2012ge}
 Baccetti, V.; Martin-Moruno, P.; Visser, M.
 Gordon and Kerr-Schild ansatze in massive and bimetric gravity.
 \emph{J. High Energy Phys.} {\bf 2012},  doi:10.1007/JHEP08(2012)108.

\bibitem{Baccetti:2012re}
 Baccetti, V.; Martin-Moruno, P.; Visser, M.
 Null Energy Condition violations in bimetric gravity.
 \emph{J. High Energy Phys.} {\bf 2012},  doi:10.1007/JHEP08(2012)148.

\bibitem{Baccetti:2012bk}
 Baccetti, V.; Martin-Moruno, P.; Visser, M.
 Massive gravity from bimetric gravity.
 \emph{\mbox{Class. Quant. Grav.}} {\bf 2013}, \emph{30}, 015004.

\bibitem{Volkov:2012wp}
 Volkov, M.S.
 Hairy black holes in the ghost-free bigravity theory.
 \emph{Phys. Rev. D} {\bf 2012}, \emph{85}, 124043.

\bibitem{Myrzakulov:2013owa}
 Myrzakulov, R.; Shahalam, M.
 Statefinder hierarchy of bimetric and galileon models for concordance cosmology. \emph{J. Cosmol. Astropart. Phys.} \textbf{2013}, \emph{10}, 047.

\bibitem{Maeda:2013bha}
 Maeda, K.-I.; Volkov, M.S.
 Anisotropic universes in the ghost-free bigravity. 	\emph{Phys. Rev. D} \textbf{2013}, \emph{87}, 104009.

\bibitem{deRham:2010ik}
 De Rham, C.; Gabadadze, G.
 Generalization of the Fierz-Pauli Action.
 \emph{Phys. Rev. D} {\bf 2010}, \emph{82},~ 044020.

\bibitem{deRham:2010kj}
 De Rham, C.; Gabadadze, G.; Tolley, A.J.
 Resummation of Massive Gravity.
 \emph{Phys. Rev. Lett.} {\bf 2011}, \emph{106}, 231101.

 \bibitem{deRham:2014zqa}
 De Rham, C.
 Massive Gravity.
 \emph{Living Rev. Rel.} {\bf 2014},  doi:10.12942/lrr-2014-7.

\bibitem{Hassan:2011hr}
 Hassan, S.F.; Rosen, R.A.
 Resolving the Ghost Problem in nonlinear Massive Gravity.
 \emph{\mbox{Phys. Rev. Lett.}} {\bf 2012}, \emph{108}, 041101.

\bibitem{Hassan:2011tf}
 Hassan, S.F.; Rosen, R.A.; Schmidt-May, A.
 Ghost-free Massive Gravity with a General Reference Metric.
 \emph{J. High Energy Phys.} {\bf 2012}, doi:10.1007/JHEP02(2012)026.

\bibitem{Hassan:2014vja}
 Hassan, S.F.; Schmidt-May, A.; von Strauss, M.
 Particular Solutions in Bimetric Theory and Their Implications.
 \emph{Int. J. Mod. Phys. D} {\bf 2014}, \emph{23}, 1443002.

\bibitem{Akrami:2015qga}
 Akrami, Y.; Hassan, S.F.; K\"{o}nnig, F.; Schmidt-May, A.; Solomon, A.R.
 Bimetric gravity is cosmologically viable. \emph{Phys. Lett. B} \textbf{2015}, \emph{748}, 37--44.

\bibitem{Deser:2012qx}
 Deser, S.; Waldron, A.
 Acausality of Massive Gravity.
 \emph{Phys. Rev. Lett.} {\bf 2013}, \emph{110}, 111101.

\bibitem{Deser:2014hga}
 Deser, S.; Sandora, M.; Waldron, A.; Zahariade, G.
 Covariant constraints for generic massive gravity and analysis of its characteristics.
 \emph{Phys. Rev. D} \textbf{2014}, \emph{90}, 104043.

\bibitem{Deser:2013uy}
 Deser, S.; Sandora, M.; Waldron, A.
 Nonlinear Partially Massless from Massive Gravity? 	\emph{\mbox{Phys. Rev. D}} \textbf{2013}, \emph{87}, 101501.


\bibitem{deRham:2013wv}
 De Rham, C.; Hinterbichler, K.; Rosen, R.A.; Tolley, A.J. Evidence for and Obstructions to Non-Linear Partially Massless Gravity. \emph{Phys. Rev. D} \textbf{2013}, \emph{88}, 024003.


\bibitem{Higuchi:1986py}
 Higuchi, A.
 Forbidden Mass Range For Spin-2 Field Theory In De Sitter Space-time.
 \emph{\mbox{Nucl. Phys. B}} \textbf{1987}, \emph{282}, 397--436.
\bibitem{Deser:2001us}
 Deser, S.; Waldron, A.
 Partial masslessness of higher spins in (A)dS.
 \emph{Nucl. Phys. B} {\bf 2001}, \emph{607}, 577--604.

\bibitem{Francia:2008hd}
 Francia, D.; Mourad, J.; Sagnotti, A.
 (A)dS exchanges and partially-massless higher spins.
 \emph{\mbox{Nucl. Phys. B}} {\bf 2008}, \emph{804}, 383--420.

\bibitem{Joung:2012rv}
 Joung, E.; Lopez, L.; Taronna, M.
 On the cubic interactions of massive and partially-massless higher spins in (A)dS.
 \emph{J. High Energy Phys.} {\bf 2012},  doi:10.1007/JHEP07(2012)041.

\bibitem{Joung:2012hz}
 Joung, E.; Lopez, L.; Taronna, M.
 Generating functions of (partially-)massless higher-spin cubic interactions.
 \emph{J. High Energy Phys.} {\bf 2013},  doi:10.1007/JHEP01(2013)168.

\bibitem{Zinoviev:2013hac}
 Zinoviev, Y.M.
 All spin-2 cubic vertices with two derivatives. \emph{Nucl. Phys. B} \textbf{2013}, \emph{872}, 21--37.


\bibitem{Zinoviev:2006im}
 Zinoviev, Y.M.
 On massive spin 2 interactions.
 \emph{Nucl. Phys. B} {\bf 2007}, \emph{770}, 83--106.

\bibitem{Deser:2012qg}
 Deser, S.; Joung, E.; Waldron, A.
 Gravitational and self-coupling of partially massless spin 2.
 \emph{\mbox{Phys. Rev. D}} {\bf 2012}, \emph{86}, 104004.

\bibitem{Boulware:1973my}
 Boulware, D.G.; Deser, S.
 Can gravitation have a finite range?
 \emph{Phys. Rev. D} {\bf 1972}, \emph{6}, 3368--3382.
\bibitem{DeFelice:2013awa}
 De Felice, A.; Gumrukcuoglu, A.E.; Lin, C.; Mukohyama, S.
 Nonlinear stability of cosmological solutions in massive gravity. \emph{ J. Cosmol. Astropart. Phys.} \textbf{2013},
doi:10.1088/1475-7516/2013/05/035.

\bibitem{Hassan:2011vm}
 Hassan, S.F.; Rosen, R.A.
 On nonlinear Actions for Massive Gravity.
 \emph{J. High Energy Phys.} {\bf 2011},  doi:10.1007/JHEP07(2011)009.

 \bibitem{Hassan:2012qv}
 Hassan, S.F.; Schmidt-May, A.; von Strauss, M.
 Proof of Consistency of Nonlinear Massive Gravity in the
 St\'uckelberg Formulation.
 \emph{Phys. Lett. B} {\bf 2012}, \emph{715}, 335--339.

\bibitem{deRham:2012kf}
 De Rham, C.; Renaux-Petel, S.
 Massive Gravity on de Sitter and Unique Candidate for Partially Massless Gravity.
 \emph{ J. Cosmol. Astropart. Phys.} {\bf 2013},  doi:10.1088/1475-7516/2013/01/035.

\bibitem{Paulos:2012xe}
 Paulos, M.F.; Tolley, A.J.
 Massive Gravity Theories and limits of Ghost-free Bigravity models.
 \emph{\mbox{J. High Energy Phys.}} {\bf 2012},  doi:10.1007/JHEP09(2012)002.

\bibitem{Volkov:2012zb}
 Volkov, M.S.
 Exact self-accelerating cosmologies in the ghost-free massive gravity---The detailed derivation.
 \emph{Phys. Rev. D} {\bf 2012}, \emph{86}, 104022.

\bibitem{Wald:1984rg}
 Wald, R.M.
 \emph{General Relativity};
University of Chicago press: Chicago, IL, USA, 1984; p. 491
 \bibitem{Nurowski:2000cq}
 Nurowski, P.; Plebanski, J.F.
 Nonvacuum twisting type N metrics.
 \emph{Class. Quant. Grav.} {\bf 2001}, \emph{18}, 341--351.

\bibitem{Liu:2013fna}
 Liu, H.-S.; Lu, H.; Pope, C.N.; Vazquez-Poritz, J.
 Not Conformally-Einstein Metrics in Conformal Gravity. \emph{Class. Quant. Grav.} \textbf{2013}, \emph{30}, 165015.

\bibitem{Deffayet:2012zc}
 Deffayet, C.; Mourad, J.; Zahariade, G.
 A note on ``symmetric'' vielbeins in bimetric, massive, perturbative and non perturbative gravities. \emph{J. High Energy Phys.}
 \textbf{2013}, doi:10.1007/JHEP03(2013)086.

\bibitem{Hassan:2014gta}
 Hassan, S.F.; Kocic, M.; Schmidt-May, A.
 Absence of ghost in a new bimetric-matter coupling. \textbf{2014},
 arXiv:1409.1909[hep-th].
\bibitem{Hinterbichler:2012cn}
 Hinterbichler, K.; Rosen, R.A.
 Interacting Spin-2 Fields.
 \emph{J. High Energy Phys.} {\bf 2012},  doi:10.1007/JHEP07(2012)047.

\bibitem{Hassan:2012wt}
 Hassan, S.F.; Schmidt-May, A.; von Strauss, M.
 Metric Formulation of Ghost-Free Multivielbein Theory. \textbf{2012},
 arXiv:1204.5202 [hep-th].
\bibitem{Bonora:1985cq}
 Bonora, L.; Pasti, P.; Bregola, M.
 Weyl Cocycles.
 \emph{Class. Quant. Grav.} {\bf 1986}, doi:10.1088/0264-9381/3/4/018.
\bibitem{Metsaev:2010kp}
 Metsaev, R.R.
 6d conformal gravity.
 \emph{J. Phys. A }{\bf 2011}, \textit{44}, 175402.

\bibitem{Boulanger:2004zf}
 Boulanger, N.; Erdmenger, J.
 A Classification of local Weyl invariants in D=8.
 \emph{Class. Quant. Grav.} {\bf 2004}, \emph{21}, 4305--4316.

\bibitem{Deser:2013gpa}
 Deser, S.; Sandora, M.; Waldron, A.
 No consistent bimetric gravity?
 \emph{Phys.\ Rev.\ D} {\bf 2013}, \emph{88},~ 081501.

\bibitem{Joung:2014aba}
 Joung, E.; Li, W.; Taronna, M.
 No-Go Theorems for Unitary and Interacting Partially Massless Spin-Two Fields.
 \emph{Phys. Rev. Lett.} {\bf 2014}, \emph{113}, 091101.

\bibitem{Garcia-Saenz:2014cwa}
 Garcia-Saenz, S.; Rosen, R.A.
 A non-linear extension of the spin-2 partially massless symmetry. \emph{\mbox{J. High Energy Phys.}} \textbf{2015},
doi:10.1007/JHEP05(2015)042.

\bibitem{PMsym}
 Hassan, S.F.; Schmidt-May, A.; von Strauss, M.
Extended Weyl Invariance in a Bimetric Model. \textbf{2015},
In preparation.

 
  
\end{thebibliography}
\end{document}